\documentclass[aps,prb,amsmath,twocolumn,amssymb,titlepage,nofootinbib]{revtex4-2}

\usepackage{graphicx}
\usepackage{nicefrac}
\usepackage{amsfonts}
\usepackage{ulem}
\usepackage{amssymb}
\usepackage{amsmath} 

\usepackage{subfigure}
\usepackage{multirow} 
\usepackage{tabularx} 
\usepackage{array}

\usepackage{units}

\usepackage{tensor} 
\usepackage{braket}

\usepackage{bm}
\usepackage{hyperref}

\usepackage{colortbl}

\definecolor{SuperElNiñoColor}{rgb}{1,1,0.5} 

\begin{document}

\author{Chae-Hyun Yoon }
\author{Jubin Park }
\author{Myung-Ki Cheoun }
\affiliation{Department of Physics and Origin of Matter and Evolution of Galaxies (OMEG) Institute, Soongsil University, Seoul 06978, Republic of Korea}

\vspace{-1em}

\date{\today}

\title{Deciphering Super El Niño: Development of a Novel Predictive Model Integrating Local and Global Climatic Signals}

\begin{abstract}
In recent years, extreme weather events have surged, highlighting the urgent need for action on the climate emergency. 
The year 2023 saw record-breaking global temperatures, unprecedented heatwaves in Europe, devastating floods in Asia, 
and severe wildfires in North America and Australia, which indicate significant changes in climate patterns driven by 
natural and anthropogenic factors. Super El Niño events, known for their profound impact on global weather, 
play a critical role in these changes, causing severe economic and environmental damage. 
This study presents a novel predictive model that integrates systematically local and global climatic signals to forecast Super El Niño events, 
introducing the Super El Niño Index (SEI), which value of 80 or higher defines a Super El Niño event. 
Our analysis shows that the SEI accurately reflects past Super El Niño events, including those from 1982-83, 1997-98, 
and 2015-16, with SEI values for these periods containing 80 within the 2-sigma standard deviation. Using data up to 2022, 
our model predicted an SEI of around 80 for 2023, indicating a Super El Niño for the 2023-24 period. 
Recent observations confirm that the 2023-24 El Niño is among the five strongest recorded Super El Niño events in history. 
An analysis of SEI trends from 1982 to 2023 reveals a gradual increase, 
with recent El Niño events consistently exceeding SEI values of 70. 
This trend suggests that El Niño events are increasingly approaching Super El Niño intensity, 
potentially due to more favorable conditions in the equatorial Pacific. 
This increase in SEI values and the frequency of stronger El Niño events may be attributed to the ongoing effects of global warming. 
These findings emphasize the need for heightened preparedness and strategic planning to mitigate the impacts of future Super El Niño events, 
which are likely to become more frequent in the coming decades.
\end{abstract}%


\maketitle

\section{Introduction}
In recent years, the world has witnessed an alarming escalation in extreme weather events, 
underscoring the urgent need to address climate change. 
The year 2023 marked the highest global average temperature on record, 
surpassing previous records by a significant margin\,\cite{ScienceDaily_2024_01}. 
Unprecedented heatwaves scorched Europe, severely impacting human health and infrastructure\,\cite{ScienceDaily_2023_08}. 
Meanwhile, devastating floods inundated parts of Asia, displacing millions and causing substantial economic losses\,\cite{WMO_2024_04}. 
In 2022, record-breaking wildfires ravaged North America and Australia, 
further highlighting the severe and multifaceted impacts of climate anomalies\,\cite{Phys.org_2024_01,Copernicus_2024}. 
These events are symptomatic of a statistically significant alteration in long-term climate patterns, 
driven by both natural phenomena and anthropogenic influences. 
Since the onset of the Industrial Revolution in the 18th century, 
human activities have increasingly influenced the climate, 
leading to a sharp rise in global greenhouse gas emissions. 
According to the IPCC (2007)
\footnote{
The Intergovernmental Panel on Climate Change (IPCC), established in 1988, 
is an intergovernmental body that assesses the science of climate change. 
It produces regular assessment reports, special reports, methodology reports, and technical papers. 
The IPCC's reports are widely regarded as the most authoritative source on climate change, 
informing global climate policy and mitigation strategies.
}, these emissions surged by approximately $70\%$ from 1970 to 2004, 
with projections forewarning catastrophic consequences for our planet by 2030 if current trends continue.


The ramifications of climate change extend across various sectors, including 
agriculture, forestry, marine ecosystems, and ecological systems. 
Coastal regions are particularly vulnerable, facing heightened risks of flooding and 
intensified natural disasters such as typhoons and hurricanes due to accelerated sea level rise\,\cite{D. Breitburg_2018}. 
Similarly, inland areas grapple with challenges like 
the spread of tropical diseases, heatwaves, droughts, forest fires, and desertification\,\cite{Y. Kang_2009}. 
Despite efforts to mitigate climate change through policy interventions, 
the persistent impact of climate change is expected to endure for decades, 
emphasizing the urgency of recognizing its severity and devising strategies to minimize damage.

\begin{table*}[]
	\begin{tabular}{lll}
		Method                               & Description                                       & Note \\ \hline \hline
		\multirow{2}{*}{SST Anomalies{\,\cite{P. J. Minnett_2019_SST}}}       & Measures the average sea surface temperature      & Direct relation to El Niño. Historical dataset available. \\
		& anomalies in the central Pacific.                 & Might not capture global effects. \\ \hline
		\multirow{2}{*}{ONI{\,\cite{M. H. Glantz_2020_ONI}}}                 & A three-month running average of SST anomalies    & Consistent representation. Primary benchmark. \\
		& in the Niño 3.4 region                            & for El Niño intensity. Lag in real-time prediction. \\ \hline
		\multirow{2}{*}{SOI{\,\cite{N. Chen_2017_SOI_Dynamic}}}                & Measures the atmospheric pressure differences     & Uses atmospheric data. Long historical data. \\
		& between Tahiti and Darwin.                        & Can be influenced by other factors. \\ \hline
		\multirow{2}{*}{Dynamic Models{\,\cite{N. Chen_2017_SOI_Dynamic}}}      & Complex computational models simulating           & Holistic view. Long-term forecasts. \\
		& interactions between atmosphere, oceans, and land & Heavy computational resources needed. \\ \hline
		\multirow{3}{*}{Statistical Models{\,\cite{X. Wang_2020_static}}}  & \multirow{2}{*}{Uses historical data to identify patterns}    & Less resource-intensive. \\
		& \multirow{2}{*}{based on statistical correlations.}           & Can be accurate for repetitive patterns. \\
		&                                                               & Relies on repetitive patterns. \\ \hline
		\multirow{2}{*}{Hybrid Models{\,\cite{S. Wang_2021_hybrid}}}       & \multirow{2}{*}{Combines dynamic and statistical approaches.}  & Strengths of both dynamic and statistical models. \\ 
		&                                                   & Still in development. \\ \hline
	\end{tabular}
	\caption{Well-known methods for Predicting El Niño phenomena. 
		SST, ONI, and SOI are acronyms for sea surface temperature, the Oceanic Niño Index, and the Southern Oscillation Index, respectively.
		In particular, the advantages and limitations of each model are briefly introduced in the Note.}
	\label{tab:methos_ENs}
\end{table*}
Among the various climate phenomena influenced by global warming, 
El Niño\,\cite{EL_Trenberth_1997} stands out due to its significant impact on global weather patterns. 
El Niño is a periodic climate phenomenon characterized by 
the warming of the central and eastern tropical Pacific Ocean. 
The term "El Niño" originates from Spanish, meaning "the little boy," referring to the Christ child, 
as this warming typically occurs around Christmas. 
El Niño events generally occur every 2 to 7 years and last for about 9 months to a year. 
During an El Niño event, the trade winds weaken or even reverse direction, 
allowing warm water from the western Pacific to flow back eastward, 
thereby warming the central and eastern Pacific Ocean. 
This redistribution of heat disrupts normal weather patterns, 
leading to significant climatic and environmental impacts across the globe.

\begin{figure*}
	\centering
	\includegraphics[]{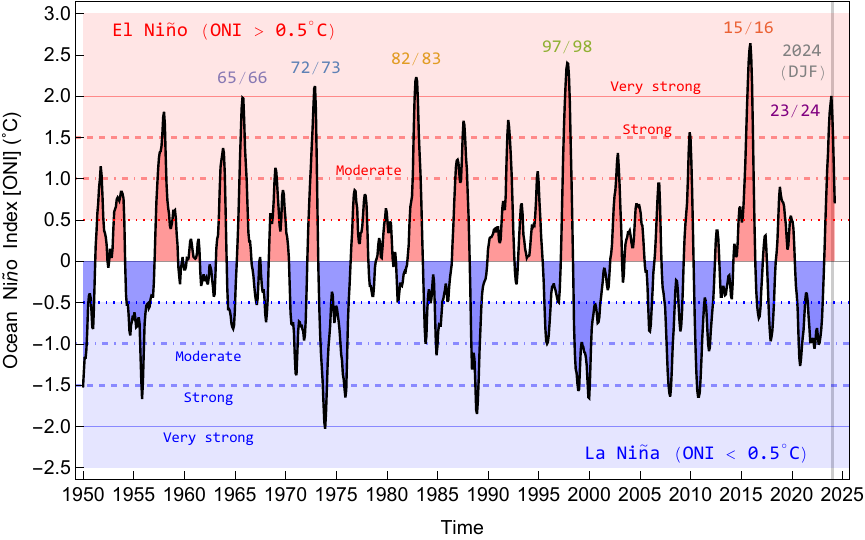}
	\caption{
ONI from January 1950 to January 2024, highlighting El Niño (red region, ONI $>$ 0.5) and La Niña (blue region, ONI $<$ -0.5) phases.
Horizontal lines represent different intensity levels: weak (dotted line, ONI=0.5), moderate (dot-dashed line, ONI=1.0), 
strong (dashed line, ONI=1.5), and very strong (solid line, ONI=2.0). 
Notably, past Super El Niño events (indicated by year labels: 65/66, 72/73, 82/83, 97/98, 15/16) all reached or 
exceeded an OSI maximum of 2.0 (peaks). 
The peak values for the Super El Niño events in 1997/98 and 2015/16 were approximately +2.4 and +2.6 degrees Celsius, respectively. 
The vertical gray line marks the recent 2023/24 period, potentially the fifth Super El Niño season on record.
	}
	\label{fig:ONI}
\end{figure*}

During an El Niño event, the trade winds weaken or even 

Table \ref{tab:methos_ENs} presents the methods used for forecasting El Niño. 
Each method has its advantages and limitations. 
Sea Surface Temperature (SST) anomalies measure central Pacific sea surface temperature deviations. 
This provides a historical dataset essential for El Niño analysis but may miss broader global impacts\,\cite{P. J. Minnett_2019_SST}. 
The Oceanic Niño Index (ONI) uses a three-month average of SST anomalies in the Niño 3.4 area. 
It offers a consistent measure of El Niño's intensity but can delay real-time forecasts\,\cite{M. H. Glantz_2020_ONI}. 
The Southern Oscillation Index (SOI) uses atmospheric pressure differences between Tahiti and Darwin, 
benefiting from extensive historical records but is affected by other variables\,\cite{N. Chen_2017_SOI_Dynamic}. 
Dynamic models simulate interactions among the atmosphere, ocean, 
and land for long-term forecasts but require significant computational resources\,\cite{N. Chen_2017_SOI_Dynamic}. 
Statistical models analyze historical data to find correlations with fewer resources 
but may lack precision due to reliance on past patterns\,\cite{X. Wang_2020_static}. 
Emerging hybrid models combine the strengths of dynamic and statistical models, 
aiming for more reliable forecasts. 
These models are still being developed but show promise in overcoming the limitations of individual methods\,\cite{S. Wang_2021_hybrid}.

Figure.\,\ref{fig:ONI} shows the ONI from January 1950 to January 2024, highlighting occurrences of El Niño and La Niña. 
To be officially recognized as an El Niño or La Niña event, 
the ONI must reach or exceed +0.5°C above average or -0.5°C below average, respectively, 
and persist for at least five consecutive, overlapping three-month period
\footnote{
	The criteria for defining El Niño and La Niña events via the ONI differ internationally. 
	In South Korea and the United States, the standard ONI threshold is $\pm ±0.5$, measured over five months. 
	This is in contrast to Australia, which uses an ONI value of $\pm 0.8$, and Japan, which considers $\pm 1.0$, both over a six-month period. 
	These variations reflect regional adaptations in monitoring and forecasting these significant climatic phenomena.
	}. 
In Figure.\,\ref{fig:ONI}, El Niño events are shown in red areas (ONI $> 0.5$), 
while La Niña events are shown in blue areas (ONI $< -0.5$).
Intensity levels are indicated by horizontal lines: 
dotted for weak events (ONI = $\pm 0.5$), dot-dashed for moderate events (ONI = $\pm 1.0$), 
dashed for strong events (ONI = $\pm 1.5$), and solid for very strong events (ONI = $\pm 2.0$).
The figure emphasizes previous Super El Niño (SE) events, 
marked with specific years (65/66, 72/73, 82/83, 97/98, 15/16), 
all of which reached or exceeded an ONI peak of 2.0, 
indicating significant intensity\,\cite{M. H. Glantz_2020_ONI, N. Chen_2017_SOI_Dynamic}.
\footnote{
The SST differences from the 30-year average 
for Super El Niños (SEs) are also shown in Fig.\,\ref{fig:DAT_Nino3.4} (in the same colors).
}
Notably, the well-documented SE events of 1997/98 and 2015/16 had 
peak values of approximately +2.4°C and +2.6°C, respectively. 
The vertical gray line marks the recent 2023/24 period, 
which was the fifth Super El Niño season on record.
Additionally, it is interesting to note that 
La Niña events have always followed SE events, 
highlighting a significant pattern in climatic behavior\,\cite{X. Wang_2020_static}.

SE events\,\cite{SEL_McPhaden_97-98}, characterized by exceptional warmth in the Pacific Ocean, 
stand out from standard El Niño events in both intensity and impact. 
Notable SE events, such as those in 1997-98 and 2015-16, saw the Niño 3.4 index reaching values 
exceeding +2.4°C and +2.6°C, respectively, significantly higher than the standard El Niño threshold of +0.5°C. 
Although there is no universally agreed-upon definition for a SE, 
the intensity of sea surface temperature anomalies is a key factor. 
For our classification, we consider El Niño events with Niño 3.4 index values exceeding +2.0°C as SE events. 
Unlike regular El Niño events, which can alter global climate patterns, 
SE events unleash far more extreme weather phenomena, profoundly impacting worldwide weather systems, 
economies, and ecosystems. Predicting SE events is critically important, 
as advance warning can help mitigate potentially catastrophic outcomes.

Forecasting a Super El Niño is particularly challenging and differs from 
predicting a regular El Niño due to the extreme anomalies involved. 
These events are rare and their prediction requires precise modeling that 
can capture the complex oceanic and atmospheric interactions on a larger scale and intensity.
Compared to typical El Niño events, which occur at irregular intervals every two to seven years, 
Super El Niños are less frequent but significantly more intense, 
often associated with the strongest phase of the El Niño cycle. 
The distinctive features and occurrence patterns of Super El Niños demand careful monitoring 
and advanced analytical models for accurate prediction and preparedness strategies. 
This sets them apart from their less intense counterparts. 

\begin{figure*}[th!]
	\centering
	\includegraphics[]{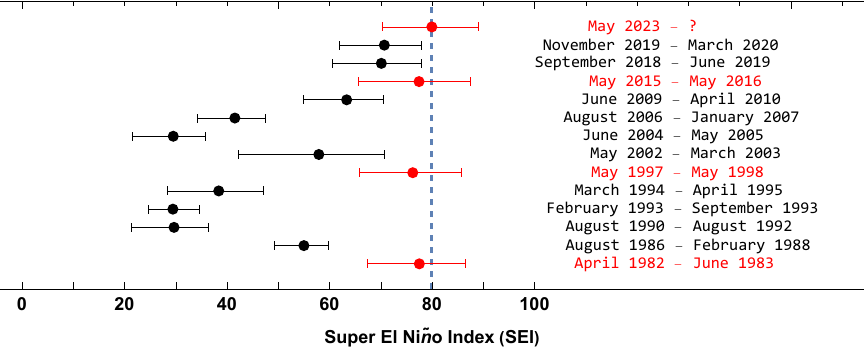}
	\caption{
Super El Niño Index (SEI) values for all known El Niño events since 1982. 
The bullet in the center of each line represents the median SEI value, 
and the corresponding fences indicate the $2\sigma$ uncertainty of this value. 
The vertical blue dotted line at SEI=80 represents the threshold distinguishing Super El Niño from regular El Niño. 
The historic Super El Niño events of 1982-83, 1997-98, and 2015-16 are highlighted in red. 
Note the decreasing difference between the SEI values of regular El Niño and Super El Niño events over time since 1982. 
Additionally, the SEI values of El Niño events in 2018-19 and 2019-20 are close to the Super El Niño threshold (SEI=80). 
Recently, a high SEI value around 80 was predicted since May 2023, 
and recent observations confirmed that this El Niño corresponds to a Super El Niño.
	}
	\label{fig:SEI}
\end{figure*}

In this comprehensive study, we present a new SE prediction model. 
The model incorporates temperature data from the El Niño monitoring area in the Pacific, 
sea surface temperature and height around the Korean Peninsula, and 
data from cities representing the three sides of the Korean Peninsula. 
This approach highlights the significance of remote correlations in SE events and the influence of distant continents.

Our careful correlation analysis reveals that SE events can impact regions far beyond the equatorial zone, 
extending as far as the Korean Peninsula in the Northern Hemisphere. 
During SE events, the warm ocean currents in the East/Central Pacific travel along the North Equatorial Current to the Philippine Sea, 
then move north along eastern Taiwan into the Kuroshio Current, significantly affecting Korea and Japan. 
Branches of this current flow into the South Sea between Korea and Japan, with some parts entering the East Sea (Sea of Japan) 
and others returning to the West Sea (Yellow Sea). 
Notably, the ocean currents that reach the West Sea have a substantial impact on winter temperatures in Seoul, 
located at the center of the Korean Peninsula. 
These findings suggest that the influence of SE events extends beyond the Pacific Ocean, 
potentially having more widespread global impacts. 
This challenges the limitations of traditional equator-centric forecasting methodologies.

We designed our model to yield the Super El Niño Index (SEI), which ranges from 0 to 100, 
as a measure of the occurrence of SE. We classify an El Niño event as a Super El Niño if its SEI value exceeds 80. 
Using this model, we identified notable SEI peaks 
during the historic Super El Niño events of 1982-83, 1997-98, and 2015-16. 
The SEI values for all known El Niño events since 1982 are shown in Fig.\,\ref{fig:SEI}.
Interestingly, 
the difference in SEI values between El Niño and SE events has been decreasing. 
This gap has significantly narrowed in El Niño events occurring after 2015, resulting in SEI values 
near to those of Super El Niño events. 
The trend of consistently high SEI values exceeding 70 since 2018 indicates that 
the environmental conditions conducive to Super El Niño events are becoming more active. 
Additionally, the linear trend line of SEI values has been gradually increasing since 1982, and by 2024, 
it is projected to increase by about 10 points (see the Fig.\,\ref{fig:SEI2}). 
This underscores the real-world impact of global warming and 
highlights the urgent need for global preparedness and adaptive strategies 
to handle more frequent Super El Niño events in the near future. 
Notably, the high SEI value around 80 in May 2023 signaled an imminent Super El Niño event, 
which recent weather observations have confirmed (For more details, see Section\,\ref{se6}).

Before concluding the introduction, the main results of this study are briefly summarized as follows:

%
\begin{itemize}
\item A new model and prediction index (SEI) have been developed to predict the SE event using data prior to 2023.
\item A new standard (SEI=80) was proposed to distinguish between El Niño (SEI $<80$)  and Super El Niño (SEI $\ge 80$).
\item Based on ONI, in the previous three Super El Niño (ONI $\ge 2.0$) cases (1982-83, 1997-98, and 2015-16), 
their SEI values reached 80 within $2\sigma$ uncertainty, 
and a sudden increase in SEI values was found to coincide with the occurrence of SE.
\item Examining the SEI for 2023, we predicted that the El Niño event had an SEI value above 80, 
and this result was confirmed by recent observations.
\item Recently, the median SEI values ($\sim 70$) of El Niños that have occurred since 2015 have approached the SE criterion (SEI$=80$) considerably. 
This reflects the fact that more environments conducive to the occurrence of Super El Niño are being created than before, 
making it easier to predict that a greater number of Super El Niño events may occur in the near future.
\item To enhance model reliability, uncertainty was incorporated. 
As a result, the uncertainty of the correlation coefficients and SEI index are shown along with their median values.
\item We demonstrate that interesting remote correlations exist between the Pacific El Niño monitoring area and 
the Korean Peninsula during Super El Niño events compared to regular El Niño events.
\item The correlation coefficients between the temperature in Seoul and 
the SST in the Niño 4, Niño 3.4, Niño 3, and Niño 1+2 regions were found to be +0.36, +0.26, 0, and -0.49, respectively. 
Interestingly, the correlation coefficient is positive in the regions closer to the Korean Peninsula and turns negative in the farthest region.
\item Similarly, the correlation coefficients between SST in the South Sea of Korea and 
the Niño 4, 3.4, 3, and 1+2 regions were 0.24, 0, -0.46, and -0.79, respectively. 
Unlike the previous correlation of 0 with the Niño 3 region for Seoul's temperature, the Niño 3.4 region also showed a correlation of 0.
\item In constructing a model to predict a Super El Niño, we used data from Niño 1+2 and Niño 4, 
both of which are correlated with Niño 3.4, the standard criterion for identifying El Niño events, 
as well as with the temperature in Seoul and the SST in the South Sea. For more detailed information, 
please refer to Fig.\,\ref{fig:15_all_coefficients}, \,\ref{fig:16_all_correlations} and \,\ref{fig:Seoul_Temp_Nino_deviation_normalized}.
\item In contrast to ONI referenced in Fig.\,\ref{fig:ONI}, SEI can predict 
whether an El Niño will develop into a Super El Niño one month prior to its occurrence.
\end{itemize}

This paper is structured as follows. 
Section \uppercase\expandafter{\romannumeral2} introduces the results of simulation studies to qualitatively support the existence of remote correlations 
and briefly discusses these results. 
Section \uppercase\expandafter{\romannumeral3} outlines the datasets used in developing the Super El Niño prediction model, 
including SSTs from Pacific El Niño monitoring areas, SSTs and sea height measurements near Korea, and temperature records from Seoul. 
Section \uppercase\expandafter{\romannumeral4} explores remote correlations between the traditional equatorial Pacific El Niño monitoring areas and 
distant regions around the Korean Peninsula in the Northern Hemisphere. 
Section \uppercase\expandafter{\romannumeral5} presents a new model for calculating SEI through these correlations obtained in Section \uppercase\expandafter{\romannumeral4}. 
Section \uppercase\expandafter{\romannumeral6} validates the 2023 Super El Niño event predicted by the model using the latest data. 
Finally, 
Section \uppercase\expandafter{\romannumeral7} concludes the paper by briefly summarizing the research results and carefully discussing their implications.

\section{Simulation study of surface ocean currents and temperatures}
In this study, we investigate how ocean currents originating from the equatorial Pacific Ocean can reach high-latitude regions 
such as Korea and Japan during (super) El Niño events. 
We utilize data from NASA’s 
ECCO2 (Estimating the Circulation and Climate of the Ocean, Phase II) project\footnote{
The ECCO2 project\,\cite{ECCO4_1, ECCO4_2, ECCO4_3} led by NASA's Jet Propulsion Laboratory, 
provides a high-resolution reconstruction of the global ocean and sea-ice states from 1992 to 2023. 
This dataset integrates diverse observational data using advanced data assimilation techniques 
within the MIT general circulation model (MITgcm)\,\cite{MITgcm}. 
For further details on the ECCO2 project and the use of MITgcm, 
refer to the project documentation and related publication.
}
to simulate three key aspects: 
1) the flow of ocean currents from the equatorial Pacific to the Northern Hemisphere, 
2) the temporal changes in SST near Korea and Japan and 
3) the SST anomalies of SE events.
For the first and second simulations,
daily average values from the two SE periods of 1997-98 and 2015-16 were used. 
For the third one, we used monthly average values from the period 1992 to 2022, 
excluding the two Super El Niño periods, to calculate the SST anomalies of the SE events. 
This approach allows us to understand the differences during the Super El Niño years 
by comparing them with the overall period's monthly averages.

\begin{figure*}
		\subfigure[Niño index regions for El Niño monitoring near the equator]
		{\label{fig:a}\includegraphics[width=11cm,height=9cm]{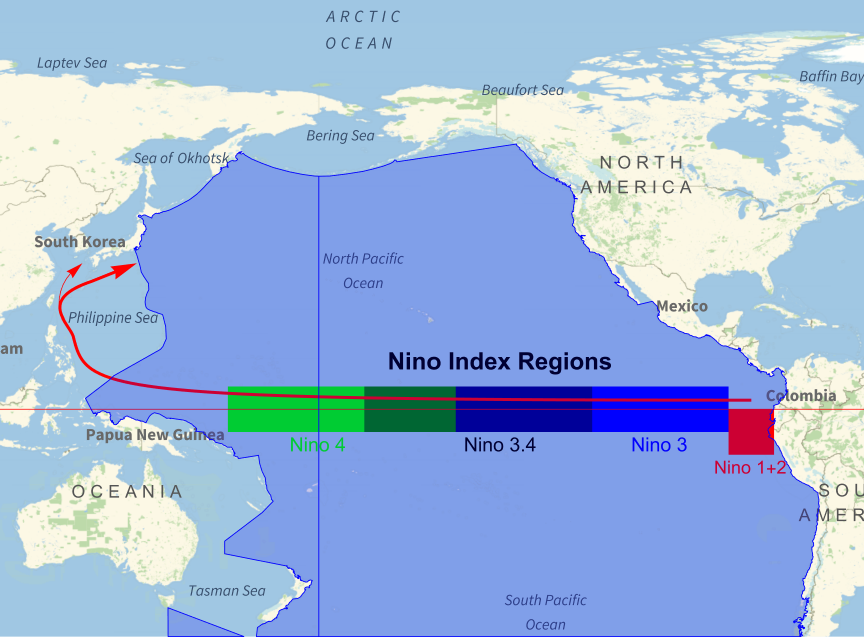}}
		\subfigure[Major ocean currents off the coasts of Korea and Japan]
		{\label{fig:b}\includegraphics[width=11cm,height=9cm]{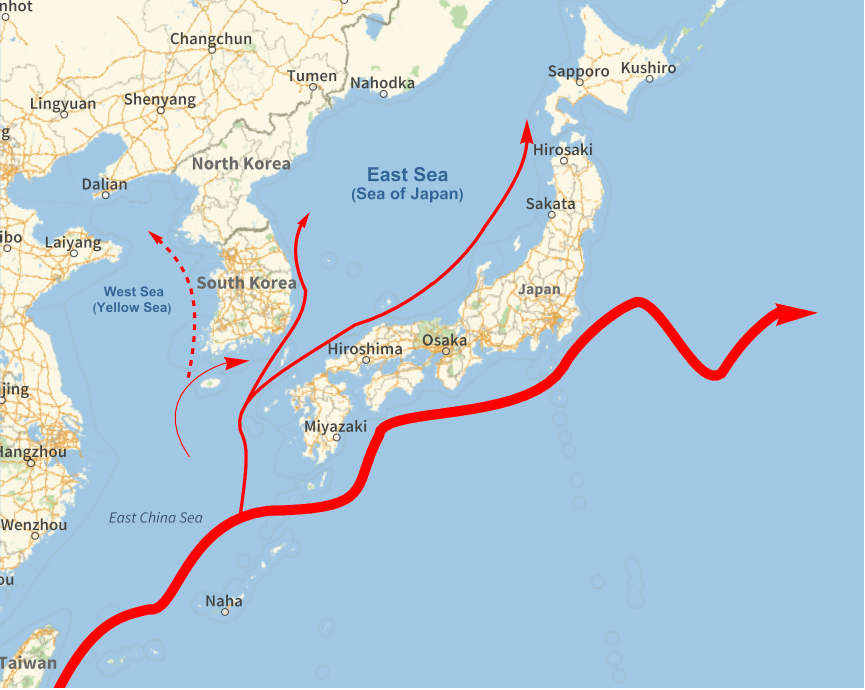}}
	\caption{Ocean currents (red arrows) related to the data used in the model. 
		\textbf{Top panel} : it shows Niño index regions near the equator in different colors. 
		The light green, dark green plus cyan, blue and red areas correspond to Niño 4, 3.4, 3, 
		and 1+2 regions, respectively.
		The thick red arrow from the equator through the Philippine Sea to the East Sea (Sea of Japan) 
		indicates the major ocean current that originate and flow near the equator.
		Additionally, a thin arrow split from the Philippine Sea show a smaller ocean current 
		flowing toward the sea between Korea and Japan.
		\textbf{Bottom panel} : it shows more specific ocean currents flowing into waters off Korea and Japan.
		Here, the thick line represents the flow of the main warm ocean current 
		called the Kuroshio Current, 
		and the thin lines represent the strength and flow of ocean currents that branch off 
		from the Kuroshio Current and flows into the enter the Korean coast 
		depending on their thickness.
		Interestingly, these split ocean currents flow less into the South and West Seas 
		than into the East Sea of Korea, 
		and note that the dashed line is used to indicate that 
		the influence of the ocean current is generally small, especially in the West Sea.
		However, if Super El Niño occurs beyond the usual El Niño, 
		the effect of Super El Niño extends to the West Sea and affects the winter temperature 
		in Seoul, a metropolitan city located on the west of the Korean peninsula.}
	\label{fig:NIR}
\end{figure*}

\subsection{Surface Ocean Currents}
\begin{figure*}
	\centering
	\subfigure[Summer (June 1st)]{\includegraphics[width=8.5cm, height=5.5cm]{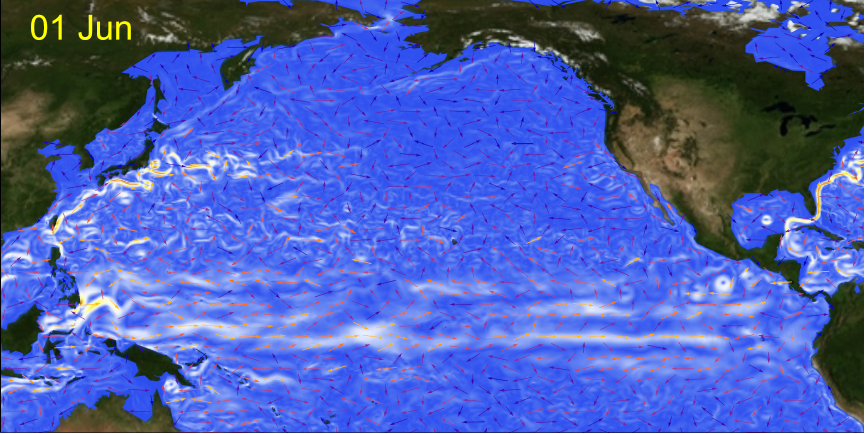}}
	\subfigure[Fall (September 1st)]{\includegraphics[width=8.5cm, height=5.5cm]{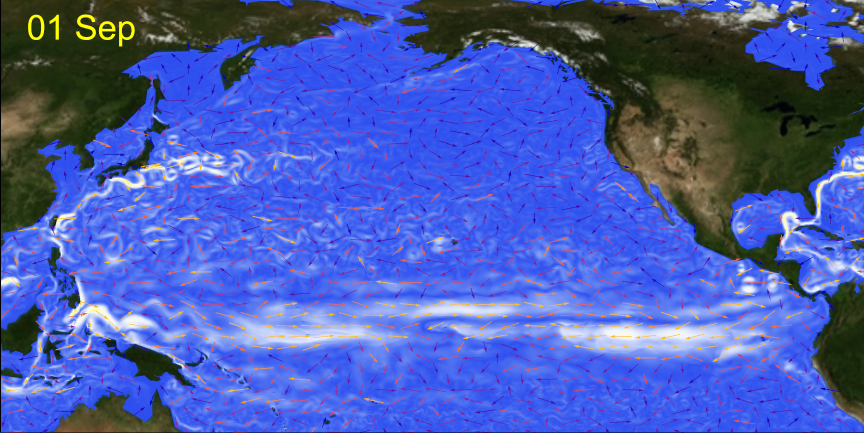}}
	\subfigure[Winter (December 1st)]{\includegraphics[width=8.5cm, height=5.5cm]{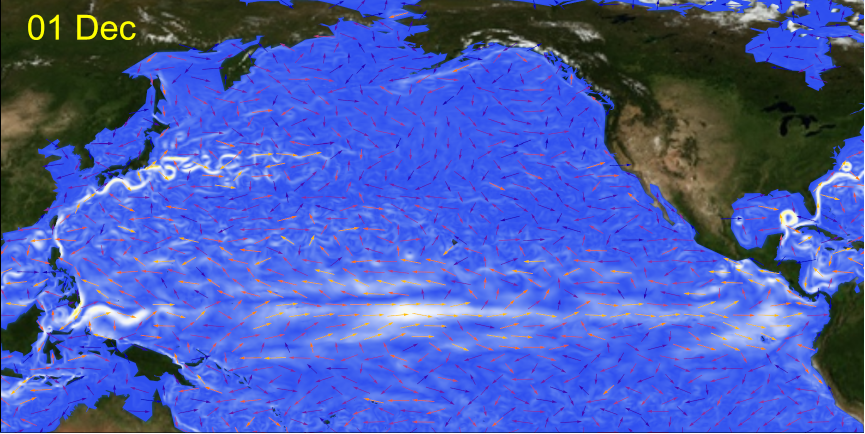}}
	\subfigure[Spring (March 1st)]{\includegraphics[width=8.5cm, height=5.5cm]{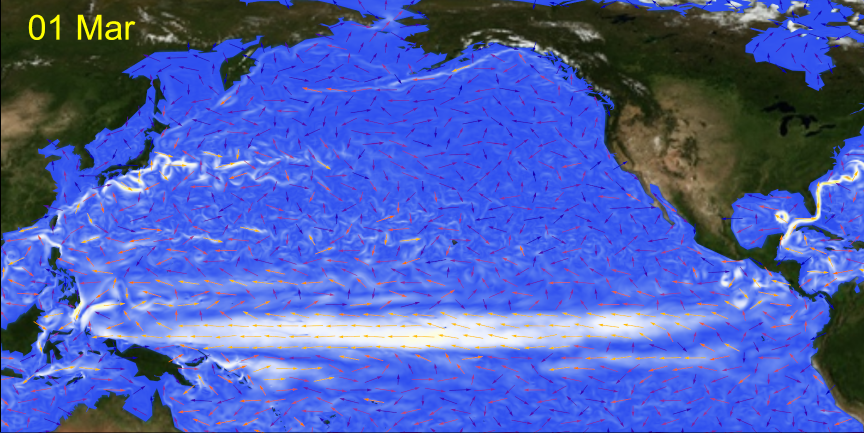}}
	\caption{Ocean flows simulated using data from NASA's ECCO2 project\,\cite{ECCO4_1, ECCO4_2, ECCO4_3}.
		From the upper left panel, it shows the flow of the North Equatorial Current 
		in summer (upper left, June), fall (upper right, September), winter (lower left, December)and spring (lower right, March).
		Colors from white to yellow-orange indicate the strength of the ocean current.
		At the bottom of each panel, the main North Equatorial Current flowing from 
		the Americas (east) to the Asia continent (west) is visible throughout the four seasons, 
		and this current (thick red arrow in the upper panel of Fig.\ref{fig:NIR}) flows north 
		from the east of western Taiwan Island, then eastward through Japan.
		In the case of fall (lower left panel), warm currents, especially near the equator, 
		can be seen moving rapidly west.
		In addition, the Kuroshio Current, which flows rapidly east through Japan, is identified, 
		and later the intensity of the flow weakens, disperses, and moves weakly east.
		In other words, at this latitude, the North Pacific current moving from west to east can be 
		confirmed as a result of this simulation.
	}
	\label{fig:OC}
\end{figure*}

The main ocean currents of interest in this study are shown as red lines in Fig.\,\ref{fig:NIR}. 
The well-known El Niño monitoring regions—Niño 1+2, Niño 3, Niño 4, and Niño 3.4—are indicated 
in red, blue, green, and bold color respectively. 
Niño 1+2, positioned near the coastline of South America, and Niño 3 and Niño 4, 
covering the eastern and central equatorial Pacific respectively, 
are vital for detecting sea surface temperature (SST) anomalies. 
Niño 3.4, which spans the central equatorial Pacific, is particularly significant for its consistent and 
robust signal of El Niño-Southern Oscillation (ENSO) variability. 
It is located between 5°S to 5°N latitude and 170°W to 120°W longitude, 
capturing variations in SSTs and atmospheric pressure anomalies. 
Due to its central location and comprehensive capture of ENSO-related variability, 
the Niño 3.4 region is essential for ENSO monitoring and prediction.

The major ocean current that originates near the equator, passes through the Philippine Sea, 
and exits east of Japan is schematically shown with a thick red arrow in the top panel of Fig.\,\ref{fig:NIR}. 
After passing Taiwan, a smaller ocean current diverges toward the sea between Korea and Japan, 
as indicated by a thin arrow branching off from the thick red arrow.

The bottom panel provides a closer look at specific currents affecting Korea and Japan. 
Here, the well-known Kuroshio Current is represented by a thick line, marking the flow of the main warm ocean current, 
while thinner lines illustrate the weakened strengths and directions of currents branching from 
the Kuroshio Current towards the Korean coast.
The ocean currents prefer to flow out to the East Sea through the South Sea of Korea. 
The current heading toward the West Sea is much weaker in strength and 
is indicated with a dotted line rather than the usual solid line to show this.
However, if a Super El Niño event occurs, surpassing the typical El Niño phenomenon, 
the influence of the warm ocean current flowing into the West Sea will be intensified, 
potentially leading to an increase in winter temperatures in Seoul, 
located close to the West Sea of the Korean Peninsula.

Geographically, the Kuroshio Current starts from the east of Taiwan in the western Pacific Ocean and flows north through Japan. 
It meets the Oyashio Current east of Japan, becoming the North Pacific Current heading east. 
Parts of the current flow into the East Sea, forming the Tsushima Warm Current and the East Korean Warm Current.

Figure\,\ref{fig:OC} shows the actual simulation results (for the currents discussed in Fig.\,\ref{fig:NIR}), 
from summer (top left panel, March first) through fall (top right panel, June first), winter (bottom left panel, September first), 
and spring (bottom right panel, December first). 
For the complete simulation results over one year, refer to the supplementary materials. 
A color scheme from white to yellow-orange indicates the strength of the currents, 
showcasing the journey of the North Equatorial Current from the Americas to Asia. 
It consistently flows north past Taiwan and then east through Japan in all seasons, 
illustrating the stability of these currents over time.

During the summer, the development of ocean currents near the southeastern equator appears as a large, 
white cloud-like shape. In the fall, these developed currents, covering the entire equatorial ocean as white clouds, 
move rapidly from west to east under the influence of the trade winds.
Additionally, strong currents generated near the equator during the fall briefly form eddies.
The Kuroshio Current, renowned for its rapid eastward flow along Japan’s coast, 
exhibits a progressive decline in intensity, transitioning into a more diffuse and sluggish eastward current.

This detailed simulation reveals the important flow of North Pacific ocean currents and complex ocean circulation patterns. 
These observations enable us to quantitatively understand the circulation of North Pacific currents, 
encompassing equatorial currents and the high-latitude Kuroshio Current. 
This understanding unveils the potential for global scale correlations, 
providing valuable insights into the intricate dynamics of the North Pacific Ocean.

\subsection{Sea Surface Temperature (SST)} 
\begin{figure*}
    \centering
    \subfigure[Summer (June 1st)]{\includegraphics[width=7.3cm, height=5.94cm]{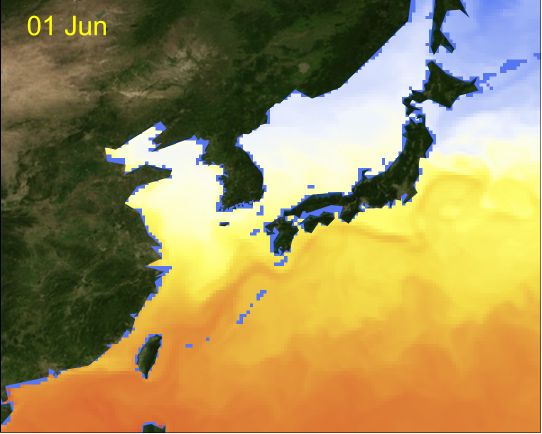}}
    \subfigure[Fall (September 1st)]{\includegraphics[width=8.5cm, height=6cm]{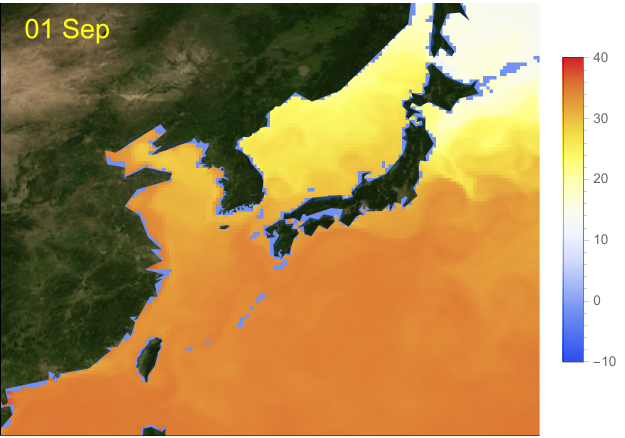}}
    \subfigure[Winter (December 1st)]{\includegraphics[width=7.3cm, height=5.94cm]{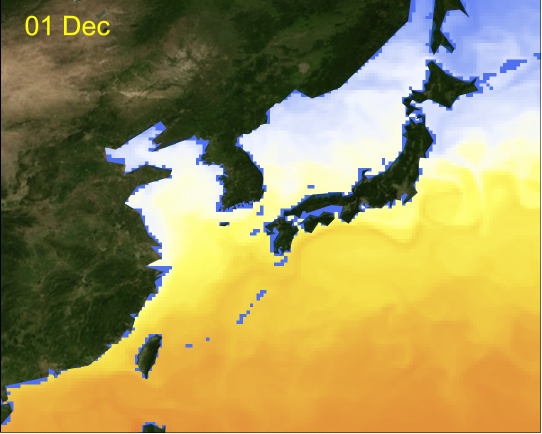}}
    \subfigure[Spring (March 1st)]{\includegraphics[width=8.5cm, height=6cm]{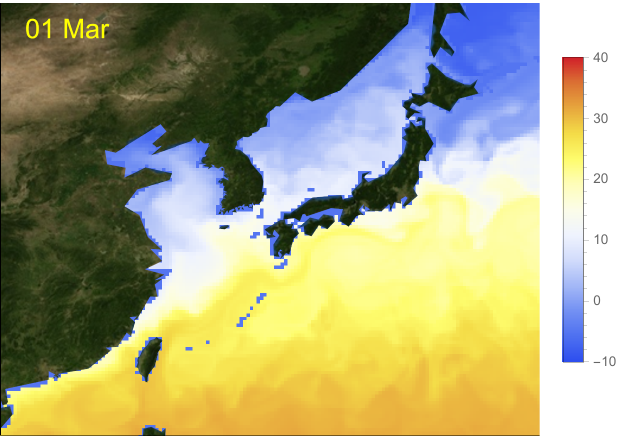}}
    \caption{As shown in Figure \ref{fig:OC}, 
    the results of simulating the sea surface temperature (SST) from summer (upper left) 
    to spring (lower right).
    The colors on each panel represent SSTs, ranging from 40 degrees (dark red) to 10 degrees below zero (dark blue).
    In spring and winter, the temperature of the East Sea and the West Sea remains cold, 
    while in summer (upper right), the hot current in the south rises, indicating that 
    the temperature of both the East and West seas rises.
    In the end, the simulation results show that in autumn, 
    all the temperatures on the west coast are above freezing.
    }
    \label{fig:SST}
\end{figure*}
In order to more qualitatively examine why remote correlations are created 
from the Pacific Ocean to the Korean Peninsula in the Northern Hemisphere, 
we simulated not only ocean currents but also the ocean surface temperature around the Korean Peninsula. 
In other words, we checked how the warm currents that rise along the ocean currents move according to the seasons.

Figure\,\ref{fig:SST} presents the results of the simulation of SST 
from summer (upper left panel) to next spring (lower right panel), 
showcasing the temperature range from 40 degrees (dark red) above zero (white) to 10 degrees (dark blue) below zero. 
It reveals that during spring and winter, the SSTs of both the East and West Seas around Korea remain low (blue color). 
In contrast, in summer (upper left panel) and fall (upper right panel), 
warm currents from the south cause the temperatures of these seas to increase, 
as evidenced by the warmer colors on the map. 
The east and west seas of Korea have changed from predominantly white to yellow, indicating the warming.

In particular, hot currents along the Kuroshio Current begin to rise 
(from south to north) along Korea's west and east seas in summer, intensify in fall, 
and then turn to sub-zero temperatures in winter (due to the descent of cold currents from the north). 
Interestingly, in this SST simulation, we can also track and 
see the year-round Kuroshio Current flow as the darker red flow in each panel.

These simulations reveal that the SE event can extend its influence beyond the equator, 
naturally impacting even the distant Northern Hemisphere. 
The North Equatorial Current directly affects Southeast Asia, and the Kuroshio Current, branching off from it, 
carries this influence northward to Korea and Japan. 
This suggests that the equatorial ocean state, after sufficient time for the currents to reach the Northern Hemisphere, 
can influence the marine environment around Korea and Seoul's temperature. 
This helps us understand how the extent and intensity of the SE phenomenon can affect different regions and the world.
(For the full year-round simulation results, please refer to the supplementary materials.)
\begin{figure*}
	\centering
	\subfigure[SON (September-October-November)]{\includegraphics[]{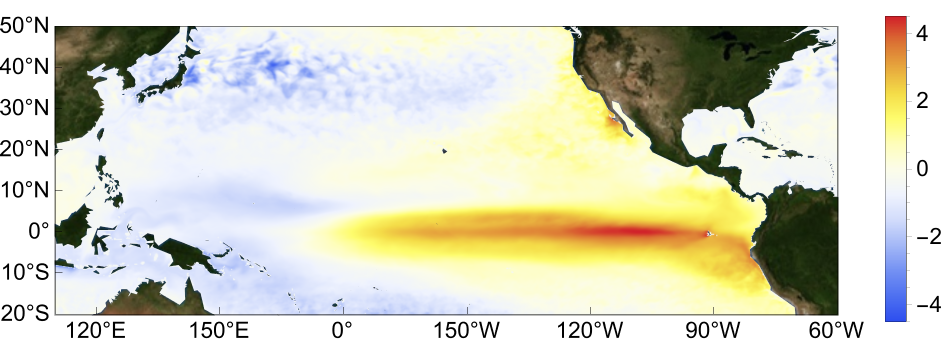}}
	\subfigure[DJF (December-January-February)]{\includegraphics[]{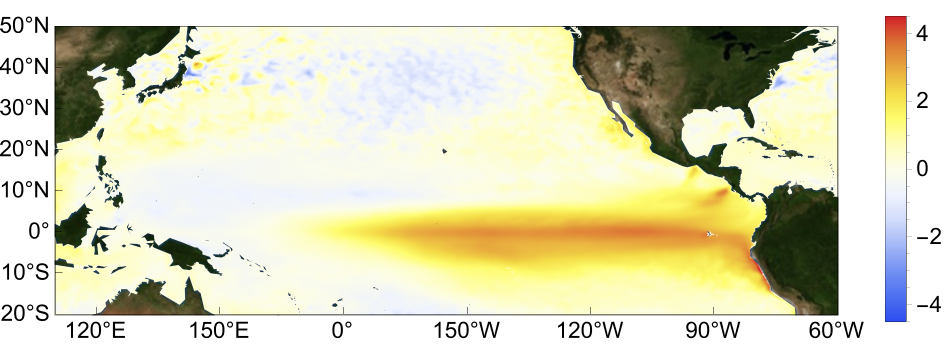}}
	\subfigure[MAM (March-April-May)]{\includegraphics[]{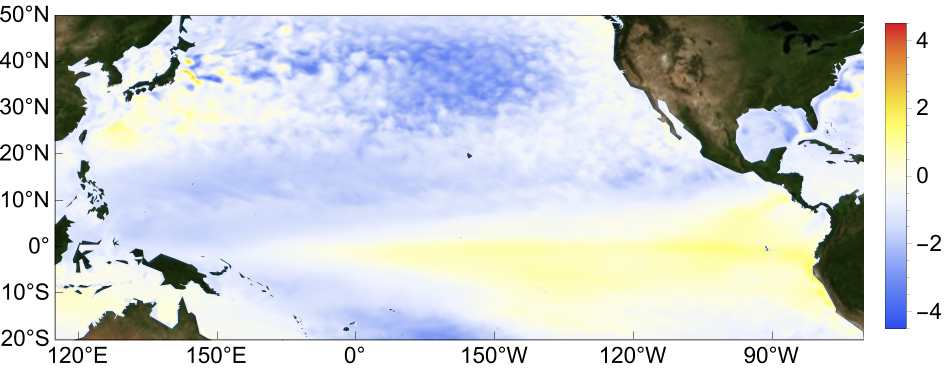}}
	\caption{3-month average SST anomalies according to the formation period (SON) - peak period (DJF) - extinction period (MAM) of Super El Niño.		
		(a) SON (September-October-November) for 30 years from 1992: SST anomalies show significant warming in the central and eastern equatorial Pacific, with temperatures reaching up to $4\,^{\circ}\mathrm{C}$ above average. 
		This period marks the onset of El Niño conditions.
		(b) DJF (December-January-February): The warming intensifies and spreads further, peaking during the winter months with anomalies exceeding $4\,^{\circ}\mathrm{C}$ in the central equatorial Pacific. This is the peak phase of the Super El Niño. Notably, temperatures near Korea and Japan also rise above average during this period, indicating the far-reaching impact of Super El Niño.
		(c) MAM (March-April-May): SST anomalies begin to weaken, although the central and eastern Pacific still exhibit warmer-than-average temperatures. The weakening of El Niño conditions is evident during this period as the SST anomalies decrease.
		These observations highlight the significant influence of Super El Niño events on distant regions, such as Korea and Japan, through mechanisms like the Kuroshio Current. This supports the concept of remote correlations, where climatic phenomena in one region can have profound effects on distant locations.
	}
	\label{fig:Delta_SST_SE}
\end{figure*}
\subsection{Analysis of Super El Niño SST Anomalies} 
In this section, we analyze the 3-month average SST anomalies during different phases of SE events. 
This analysis highlights the significant influence of SE on the central and eastern equatorial Pacific, 
as well as its far-reaching impacts on regions such as Korea and Japan.
To calculate the 3-month average SST anomalies, we utilized the ECCO2 monthly dataset spanning from 1992 to 2022. 
Monthly SST data were collected, excluding the data from the known SE periods of 1997-98 and 2015-16 
to reduce their impact on the baseline. 
Subsequently, we calculated the monthly mean SST for the entire period, establishing a reference baseline. 
For each month during the SE periods, we calculated the SST anomaly 
by subtracting the corresponding monthly mean from the observed SST. 
These monthly anomalies were then averaged over 3-month periods 
corresponding to the formation (SON; September-October-November), 
peak (DJF; December-January-February), and extinction (MAM; March-April-May) phases of SE events.

Figure 6 illustrates the results of this analysis. 
During the SON period, 
significant warming is observed in the central and eastern equatorial Pacific, with SST anomalies reaching up to 4°C above average. 
This period marks the onset of El Niño conditions, indicating the beginning of significant climatic changes in the Pacific region.

In the DJF period, the warming intensifies and spreads further, 
with anomalies exceeding 4°C in the central equatorial Pacific. 
This phase represents the peak of the SE, characterized by the most substantial temperature deviations in the central Pacific region. 
Notably, temperatures near Korea and Japan also rise above average during this period. 
This suggests that the warm anomalies are transported northward by oceanic currents such as the Kuroshio Current, 
affecting distant regions in the Northern Hemisphere and illustrating the extensive impact of SE.

During the MAM period, SST anomalies begin to weaken, 
although the central and eastern Pacific still exhibit warmer-than-average temperatures. 
This phase marks the decline of the Super El Niño, showing a gradual return to normal climatic conditions.

These observations underscore the significant influence of SE events on distant regions like Korea and Japan. 
The mechanisms driving these remote impacts include oceanic currents like the Kuroshio Current, 
which transport warm anomalies from the equatorial Pacific to higher latitudes in the Northern Hemisphere. 
This supports the concept of remote correlations, where climatic phenomena in one region can have profound effects on distant locations. 
The elevated SSTs in regions far from the equator during the peak phase of SE highlight 
the need to consider these far-reaching impacts in predictive models and climate strategies.

By analyzing these SST anomalies and understanding their implications, 
we can better predict and prepare for the wide-ranging effects of future SE events. 
This analysis lays the groundwork for developing robust predictive models 
that incorporate the complex interactions between equatorial and high-latitude climatic phenomena.

\section{Data Samples}
This section details the data samples used to develop our SE prediction model and its prediction index, SEI.
To achieve this, we use SST data\,\cite{Nino1+2, Nino3, Nino3.4, Nino4}
from traditional El Niño observation regions, 
along with another SST\,\cite{EWS_SST} data and sea surface height (SSH)\,\cite{EWS_SSH} measurements 
from specific regions in the East, West, and South Seas around the Korean Peninsula. 
Additionally, temperature data\,\cite{SeoulTemp} from Seoul is included. 
By analyzing these diverse datasets, we aim to identify intricate patterns and correlations 
within broad-scale climate events like SE. 
A comprehensive overview of each data sample follows:

\subsection{SST Data (January 1997 - December 2020) of the El Niño Observation Regions}
\begin{figure*}[ht!]
	\centering
	\includegraphics[width=16cm, height=6cm]{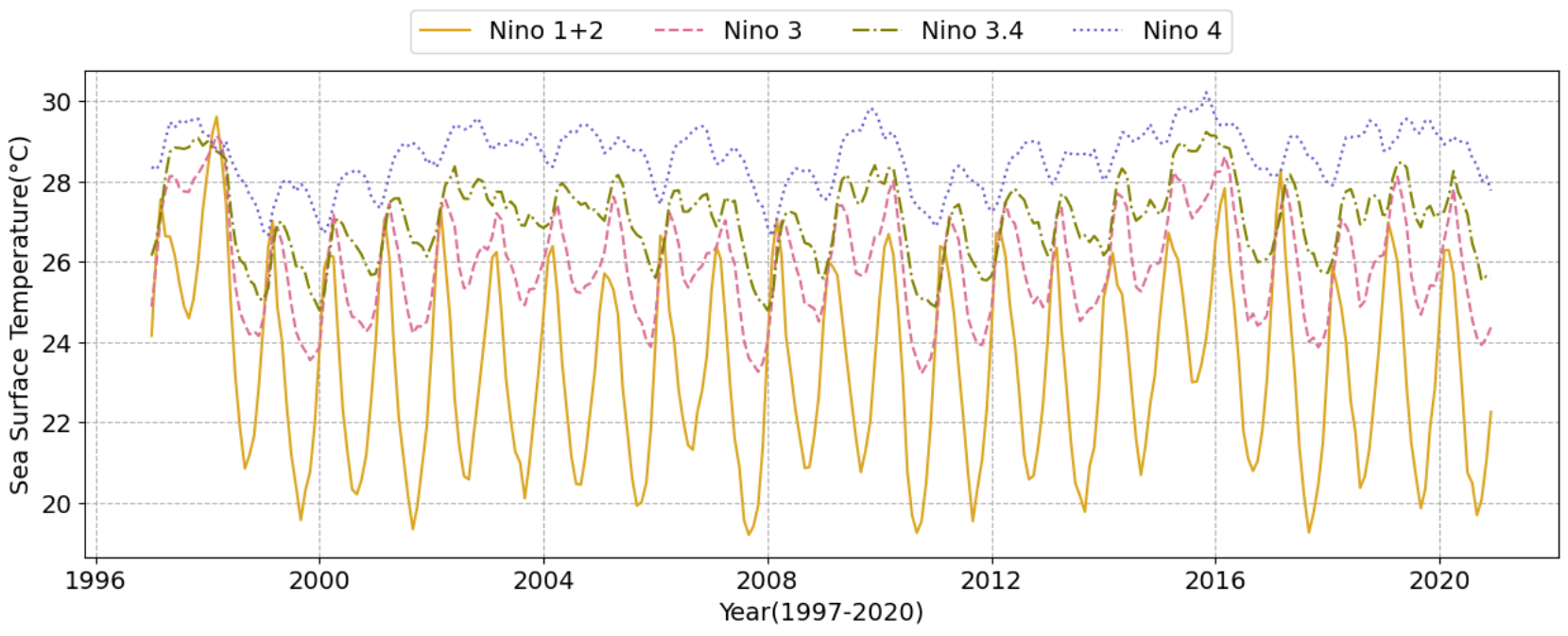}
	\caption{Sea surface temperature observations for various El Niño regions from January 1997 to December 2020. The x-axis represents time, while the y-axis indicates sea surface temperature.}
	\label{fig:6_elNino_data}
\end{figure*}
Fig.\,\ref{fig:6_elNino_data} represents the observed SST (°C) in each El Niño observation index area 
(see the upper panel in Fig.\,\ref{fig:NIR} for the location of each index area). 
Distinct color lines represent different zones: 
Niño 1+2 (brown), Niño 3 (purple dashed), Niño 3.4 (green dot-dashed), and Niño 4 (blue dotted). 
It can be seen that the Niño 1+2 area closest to the South American continent has the largest amplitude, 
and the position of the peak moves in the order of Niño 3, Niño 3.4, and Niño 4 
(according to the flow of ocean currents from east to west).

Furthermore, as it approaches the equatorial center, the amplitude of SST increases, 
leading to an overall increase in temperature. Given the seasonal variations, 
it's natural for SSTs to exhibit a yearly periodicity, a pattern that is readily observable in SST curves. 
By measuring SSTs in the Niño 3.4 region of the central Pacific and 
comparing these measurements against long-term averages, the well-known Niño 3.4 index is derived. 
This index, crucial for identifying deviations and determining SST anomalies, 
has been a key indicator of El Niño and La Niña phenomena since the 1980s. 
This era represents a leap forward in our comprehension and tracking of the ENSO, 
thanks to satellite observations and ocean buoys. 
These advancements have enabled more precise and dependable measurements of SSTs throughout the Pacific, 
enhancing our ability to monitor and understand these critical climatic events.

\begin{figure*}
	\centering
	\includegraphics[width=16cm, height=6cm]{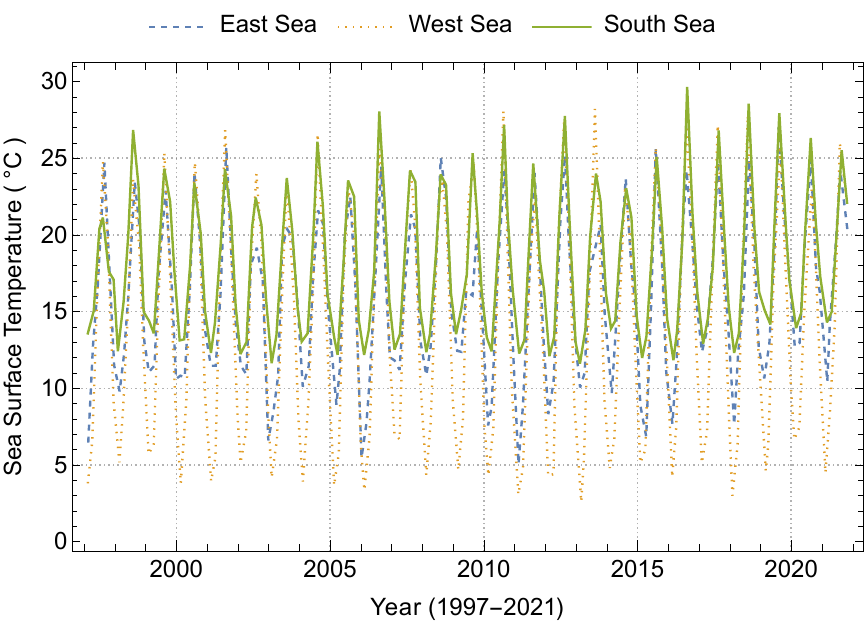}
	\caption{Sea surface temperature of the East Sea, West Sea, and South Sea in the Korean Peninsula from February 1997 to December 2021. 
		The cycle of each graph is 6 (every 2 months per year), the x-axis is time(year), and the y-axis is sea surface temperature(°C). 
		In the East Sea, the fluctuation of sea surface temperature is more changeable than the West Sea and the South Sea.}
	\label{fig:SST_EWS}
\end{figure*}

\begin{figure}
	\centering
	\includegraphics[width=8.5cm, height=8.5cm]{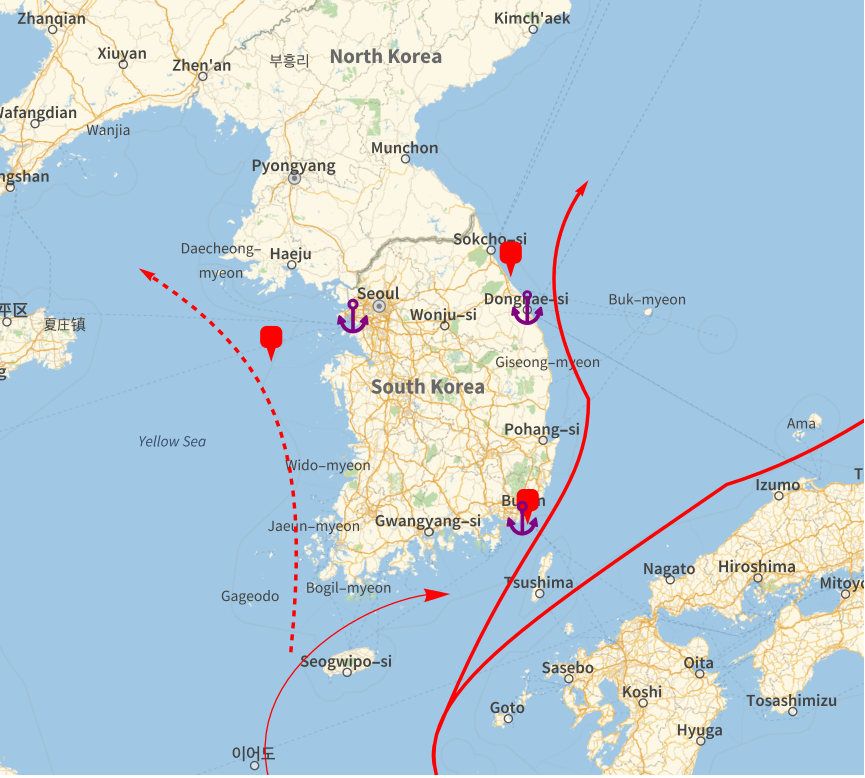}
	\caption{The positions of the East Sea, West Sea, and South Sea based on sea surface temperatures from February 1997 to December 2021. For the East Sea, the line is at 106, point at 02, and depth at 0m; for the West Sea, the line is at 307, point at 03, and depth at 0m; and for the South Sea,the line is at 207, point at 01, and depth at 0m. 
		Observation points for annual average sea surface height data: Mukho in the East Sea (blue), Incheon in the West Sea (green), and Busan in the South Sea (red).}
	\label{fig:SST_EWS_L}
\end{figure}

\subsection{SST Data (1997 - 2020) on East, West and South Seas Surrounding the Korea Peninsular} 
Fig.\,\ref{fig:SST_EWS} displays bimonthly SST readings from selected locations 
in the East Sea (represented by dashed lines), West Sea (dotted lines), and South Sea (solid lines) 
surrounding the Korean Peninsula. 
The precise locations within each sea are marked on the map of the Korean Peninsula in Fig.\,\ref{fig:SST_EWS_L}, 
using speech bubble shapes. 
Near the measurement site in the West Sea, a warm current branching off from the Kuroshio Current and 
flowing northward is shown with a thin red dotted line. 
Meanwhile, in the South Sea and East Sea, warm currents are depicted with thin and thick solid lines, respectively.
For more detailed information on the sea surface temperature in the East, West, and South Seas, please refer to Appendices B and C.

According to the simulation outcomes discussed in Section \uppercase\expandafter{\romannumeral2}, 
the intensity of the warm currents reaching the South and East Seas surpasses that of the West Sea. 
Consequently, we anticipate the highest SST readings in the South Sea, followed by intermediate values in the East Sea, 
and the lowest in the West Sea. This expectation aligns well with the SST data presented in in Fig.\,\ref{fig:SST_EWS}. 
Given that the data is collected every two months, there are six SST data points annually, 
with the lowest values recorded in February (winter in the Northern Hemisphere) and the highest in August 
(summer in the Northern Hemisphere).

\subsection{Annual Average Sea Surface Height Data (1989 - 2020)}
\begin{figure}[ht!]
	\centering
	\includegraphics[width=8cm, height=5cm]{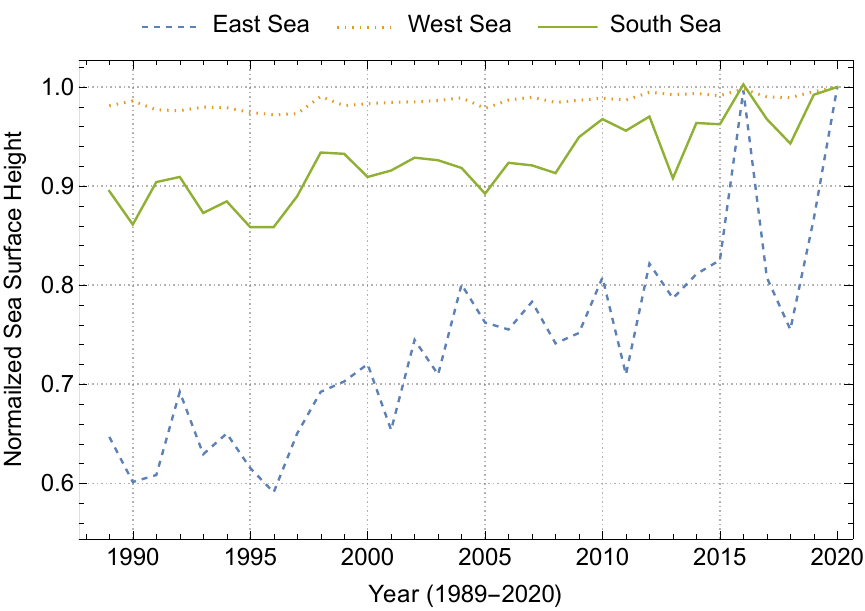}
	\caption{Annual average sea surface height (SSH) data for the East Sea, West Sea, and South Sea from 1989 to 2020. 
		The x-axis represents time (year), and the y-axis indicates sea surface height (cm). 
		Notably, all three regions show rising sea levels, with the East Sea (Mukho) experiencing an approximate 10cm increase since 1990.}
	\label{fig:annual_average_sea_surface_height}
\end{figure}
In our study, alongside SST, we also analyze annual mean sea surface height (SSH) data 
from three well-known ports on the Korean Peninsula—Mukho Port, Incheon Port, and Busan Port—
to explore the connection between (super) El Niño events and sea level changes. 
The geographical positions of these ports in the East Sea (Mukho Port), West Sea (Incheon Port), and South Sea (Busan Port) 
are depicted with anchor icons in Fig.\,\ref{fig:SST_EWS_L}. 

We normalized the sea level heights recorded 
in the East Sea (illustrated with a blue dashed line), West Sea (orange dotted line), and South Sea (green solid line) 
to their peak values for each year spanning from 1989 to 2020, as shown in Fig.\,\ref{fig:annual_average_sea_surface_height}. 
Among these, the East Sea exhibits the most significant fluctuations, followed by the South Sea and then the West Sea. 
Notably, sea levels have been on an upward trend since 1990, likely a consequence of global warming. 
The rise in sea level is particularly pronounced in the East Sea, where it has surged by approximately 10 cm since 1990, 
standing out against the more moderate increases observed in the West and South Seas.
For a detailed analysis of the rising trends and projected values of sea surface heights in the East, West, and South Seas, please refer to Appendices B and D.

\subsection{Seoul's Temperature Data (1980 - 2021)}

\begin{figure*}
	\centering
	\includegraphics[width=16cm, height=6cm]{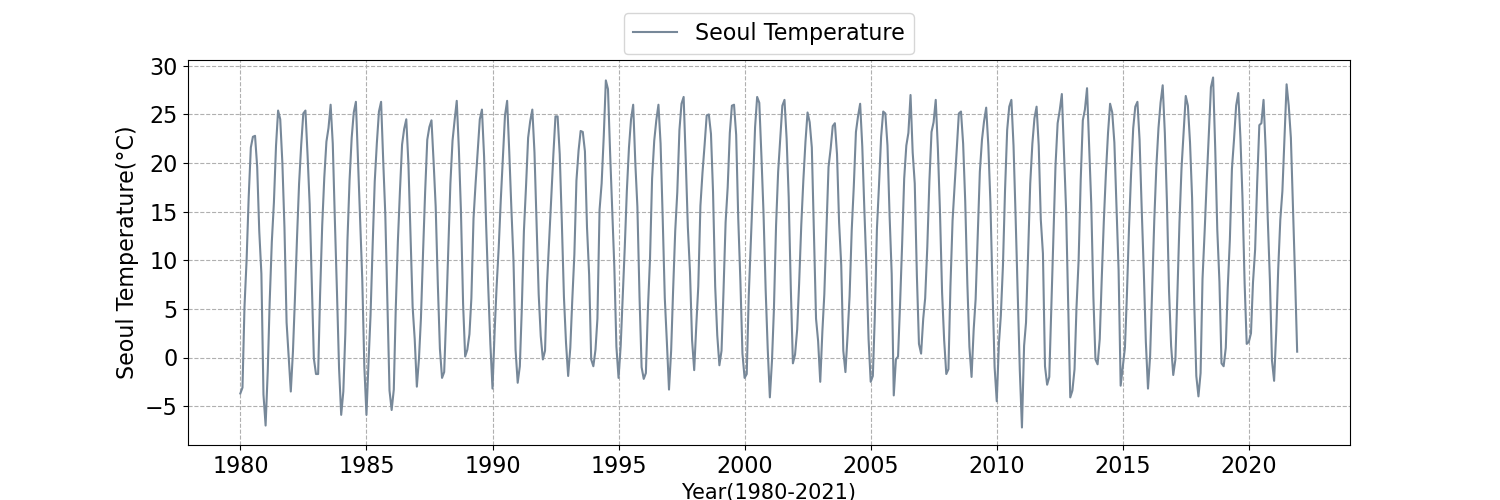}
	\caption{The average monthly temperature in Seoul from January 1980 to December 2021. 
		The x-axis is the time(year) and the y-axis is the temperature in Seoul. 
		The periodicity is 12 months.}
	\label{fig:Temp_Seoul}
\end{figure*}

Fig.\,\ref{fig:Temp_Seoul} shows the average temperature in Seoul every month from January 1980 to December 2021, 
illustrating a clear annual cycle with the lowest temperatures in January and February (Northern Hemisphere winter) 
and the highest in July and August (Northern Hemisphere summer). 
Over the observed period, temperatures range from approximately $-5\,^{\circ}$C in winter to around $25\,^{\circ}$C in summer, 
with a slight upward trend indicating a gradual increase in average temperatures over the decades. 
This trend also appears to be influenced by a variety of climatic factors, including the effects of global warming.
For detailed information on the ARIMA and BSTS analysis and forecasts regarding the temperature rise in Seoul, please refer to Appendices B and C.

\section{Correlation analysis}\label{sec.:CA}
This section presents the outcomes of a correlation analysis conducted on the diverse datasets previously described. 
The findings revealed here demonstrate that beyond the traditional Niño 3.4 region, 
there are indeed distant regions that may be sensitive to the effects of the SE phenomena. 
Furthermore, the analysis identified positive or negative correlations between specific variables.


\subsection{Correlation Analysis between Sea Surface Heights (SSH) in the East Sea, West Sea, and South Sea of the Korean Peninsula}
\begin{figure}
    \centering
    \includegraphics[width=8cm, height=5cm]{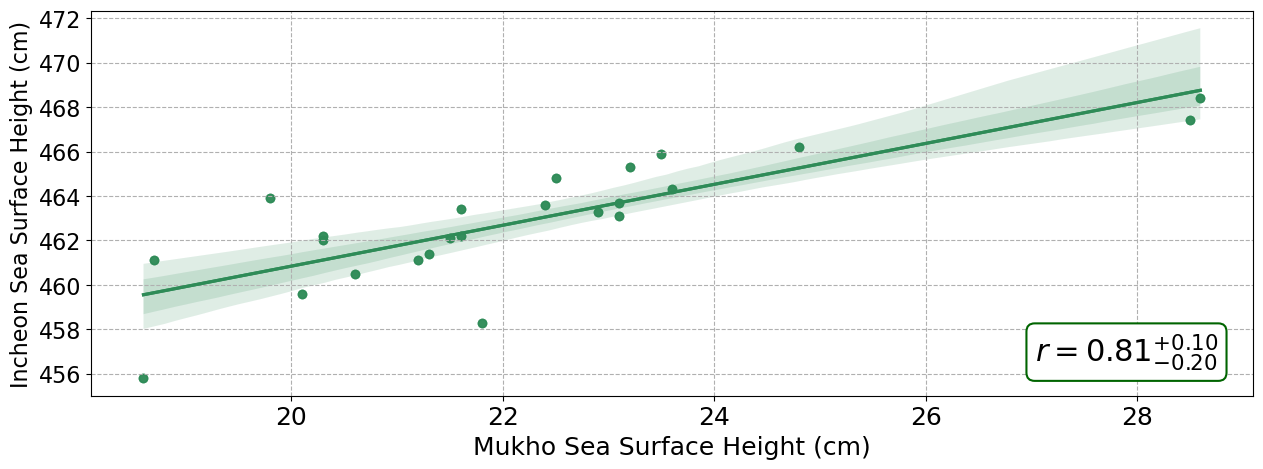}
    \includegraphics[width=8cm, height=5cm]{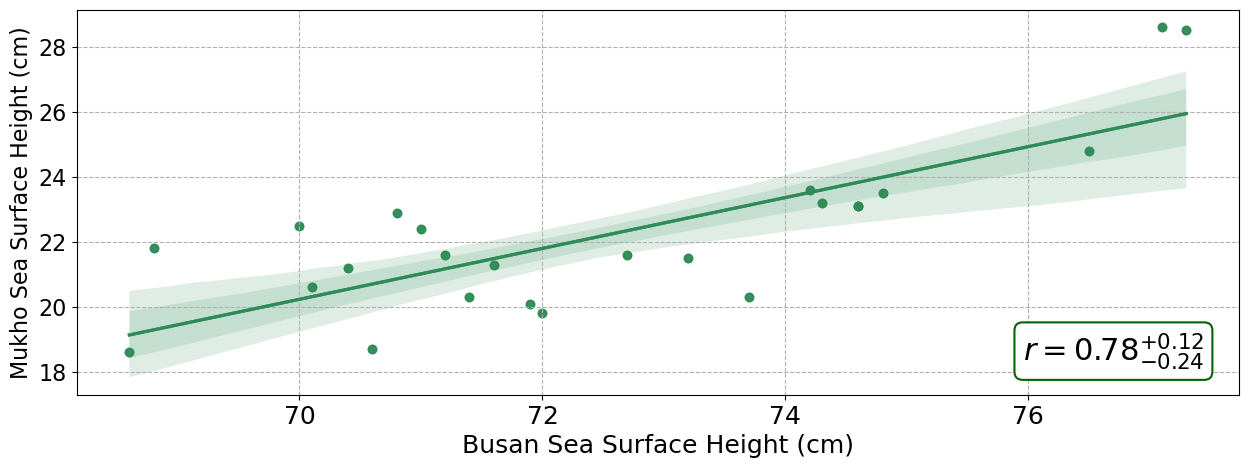}
    \includegraphics[width=8cm, height=5cm]{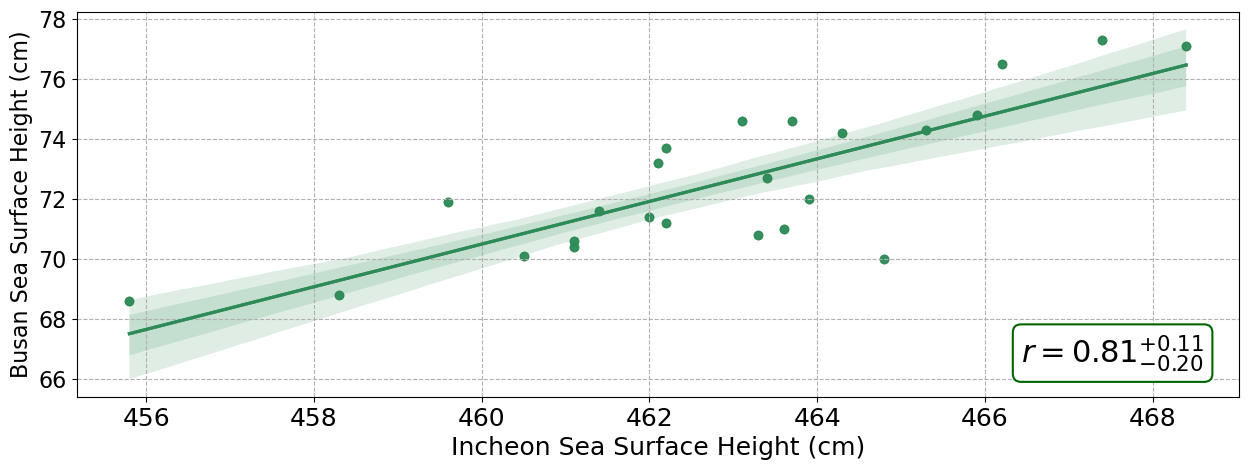}
\caption{Correlation of annual SSHs from 1997 to 2020. Top panel: Mukho (x-axis) vs. Incheon (y-axis). Middle: Busan (x-axis) vs. Mukho (y-axis). Bottom: Incheon (x-axis) vs. Busan (y-axis). In the bottom right corner of the graph, the r-values indicate the correlation coefficients for the x and y axes. The + and - values denote the 95\% confidence interval.}
    \label{fig:SSHs_CORRs}
\end{figure}

The linear correlations for these pairs, along with their $1\sigma$ and $2\sigma$ uncertainties, 
are depicted with green lines and bands, respectively, in Fig.\,\ref{fig:SSHs_CORRs}. 
The top panel shows the correlation between Mukho (x-axis) and Incheon (y-axis), 
the middle panel shows the correlation between Busan (x-axis) and Mukho (y-axis), 
and the bottom panel shows the correlation between Incheon (x-axis) and Busan (y-axis).

In our study of annual SSH data spanning from 1997 to 2020, strong positive correlations 
are found between Mukho and Incheon (0.81), Busan and Mukho (0.78), and Incheon and Busan (0.81), 
all statistically significant with the p-values below the $5\%$ level. 
These results underscore the interconnectedness of sea level changes across Korea's coastal areas. 
These high correlation values can be attributed to the fact that Mukho, Incheon, and Busan are exposed 
to the same regional sea level rise drivers, such as the Kuroshio Current, 
and are therefore influenced by common oceanic and atmospheric conditions.
The uncertainties represented by the shaded bands suggest that the correlations are robust despite the variations in data. 
The strong correlations imply that sea level changes in one area are likely to be reflected in the others.

Given the observed sea level rises in the East, West, and South Seas since 1990 and 
the strong correlations among them, 
we infer a likely common cause behind the sea level increases on the Korean Peninsula, 
potentially linked to global warming effects. 
Additionally, the consistent upward trend in sea levels, particularly in the East Sea, 
where it has surged by approximately 10 cm since 1990 (see the Fig.\,\ref{fig:annual_average_sea_surface_height}), 
highlights the regional impacts of these global phenomena.

\subsection{Analysis of the Correlation between South Sea SST and East Sea, West Sea, and South Sea SSH}
\begin{figure}
    \centering
    \includegraphics[width=8cm, height=5cm]{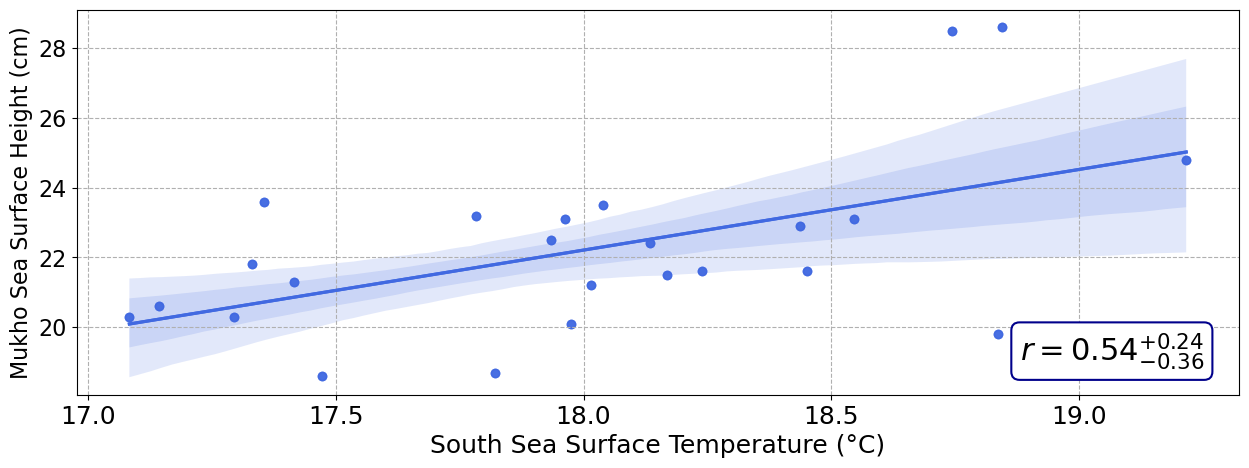}
    \includegraphics[width=8cm, height=5cm]{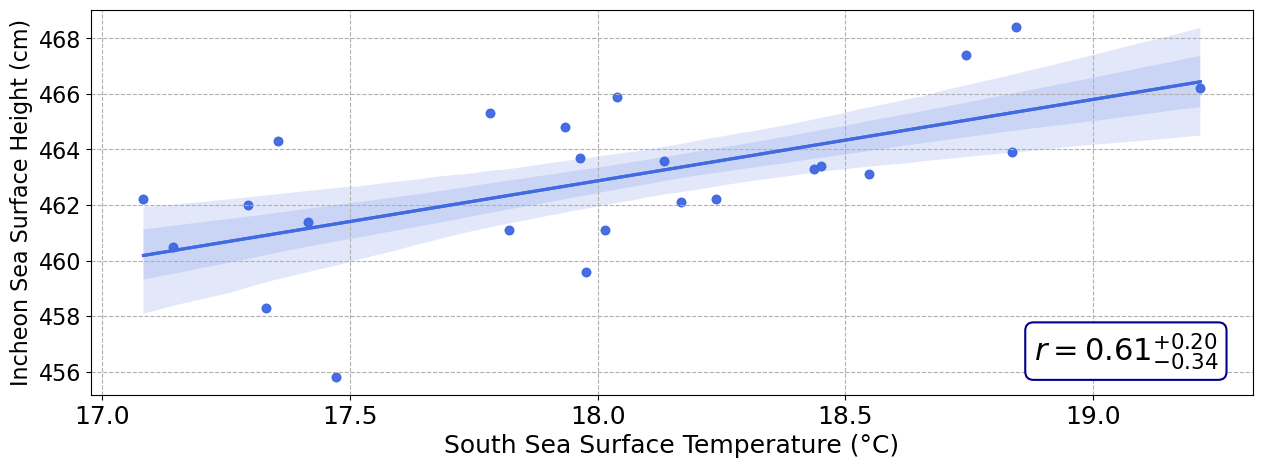}
    \includegraphics[width=8cm, height=5cm]{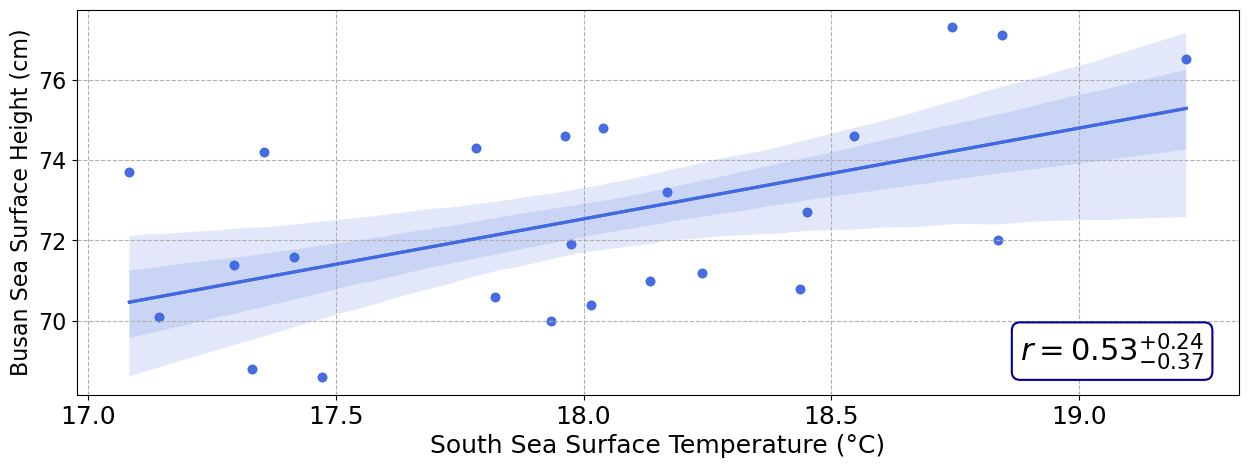}
\caption{Correlation of annual SSHs from 1997 to 2020. Top: South SST (x-axis) vs. Mukho SSH (y-axis). Middle: South SST (x-axis) vs. Incheon SSH (y-axis). Bottom: South SST (x-axis) vs. Busan SSH (y-axis).  
The meaning of the r-values with the $\pm$ values are the same as Fig. 12.}
    \label{fig:SST_SSH_CORRs}
\end{figure}

In our analysis of annual data from 1997 to 2020, we find significant correlations 
between the South Sea SST and SSHs around Korea: a positive correlation of 0.54 with Mukho Port in the East Sea, 
0.61 with Incheon Port in the West Sea, and 0.53 with Busan Port in the South Sea, 
all statistically significant with p-values below $5\%$. 
These results, illustrated in Fig.\,\ref{fig:SST_SSH_CORRs} 
with blue lines for linear correlations and blue bands for 1$\sigma$ and 2$\sigma$ uncertainties, 
indicate that SST variations in the South Sea notably influence SSHs around the Korean Peninsula. 
This suggests the thermal characteristics of the Kuroshio Current, 
flowing into the South Sea, play a critical role in regional sea level dynamics, 
highlighting the intricate link between ocean temperatures and sea level changes.

\subsection{Correlation between Annual Average Temperatures in Seoul and South Sea SST}
\begin{figure}
    \centering
    \includegraphics[width=8cm, height=5cm]{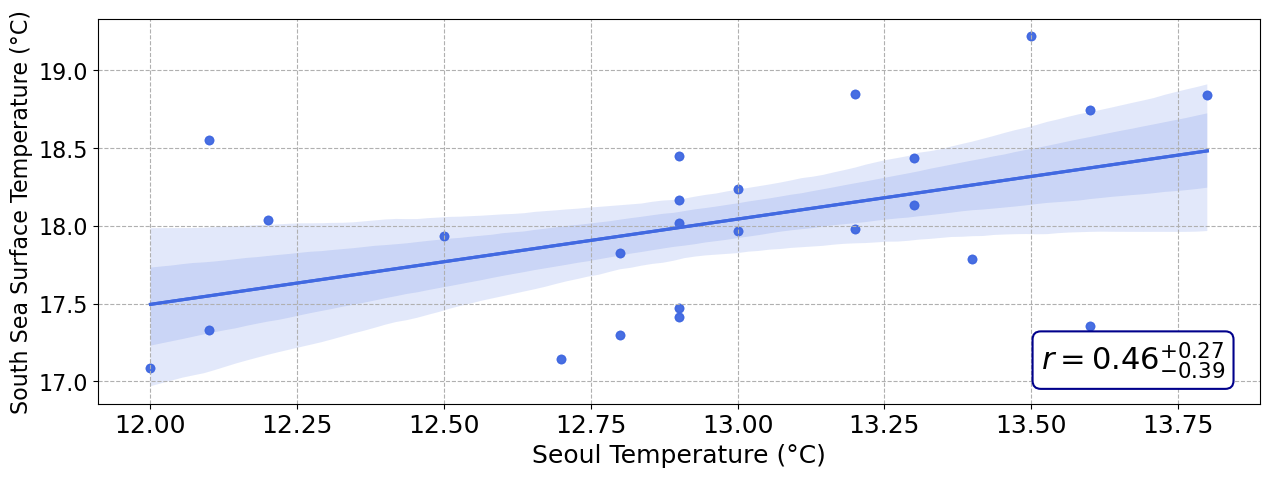}
\caption{Correlation between annual average temperatures in Seoul and South Sea surface temperatures (SST) from 1997 to 2020. The x-axis denotes Seoul's temperature, while the y-axis represents the South Sea SST. 
The meaning of the r-values with the $\pm$ values are the same as Fig. 12.}
    \label{fig:SeoulTemp_SST_CORR}
\end{figure}

Fig.\,\ref{fig:SeoulTemp_SST_CORR} depicts the correlation between Seoul's average 
temperature (x-axis) from 1997 to 2020 and the South Sea's surface temperature (y-axis), 
revealing a positive correlation of 0.46, represented by a solid blue line, 
with  1$\sigma$ and 2$\sigma$ uncertainties shown as a blue band. 
This indicates that variations in the South Sea's SST have an impact on Seoul's temperature.

\subsection{Correlation Analysis between Seoul Temperature and SSTs in El Niño Observation Areas}
Given the significant positive correlation identified between the South Sea’s annual average SST 
and Seoul’s average annual temperature, the role of the Kuroshio Current becomes evident. 
This current greatly influences the South Sea’s SST. Originating from east of Taiwan in the western Pacific, 
it flows northward past Japan into the eastern Pacific. 
The Kuroshio Current's significance is magnified during SE periods. 
During these periods, the heated North Equatorial Current shifts from the western coast of South America 
towards southern Asia, with a portion redirecting towards Taiwan. 
This backdrop sets the stage for exploring the correlation between SSTs in central and 
eastern Pacific El Niño monitoring regions and Seoul's temperatures.

\begin{figure}
    \centering
    \includegraphics[width=8cm, height=5cm]{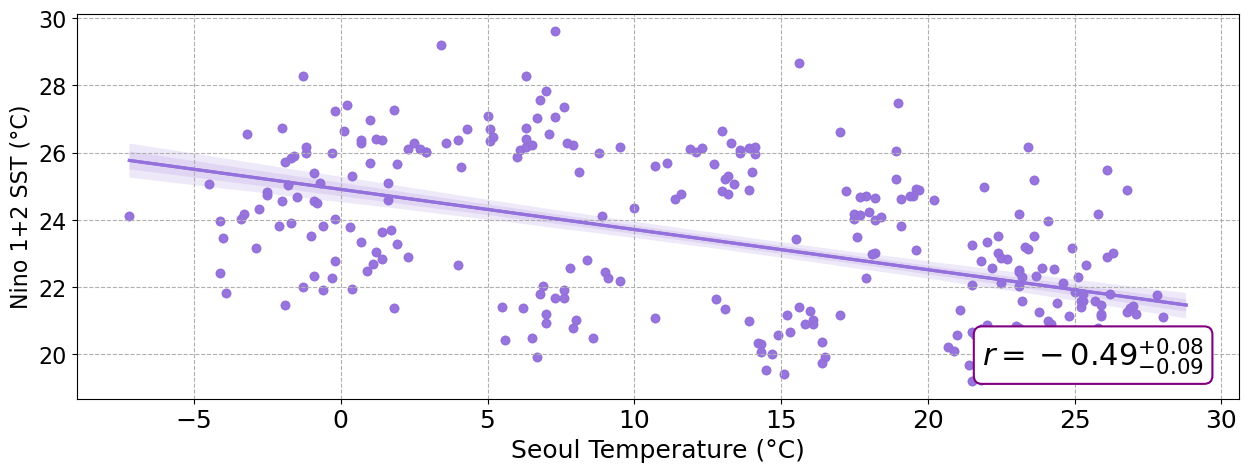}
    \includegraphics[width=8cm, height=5cm]{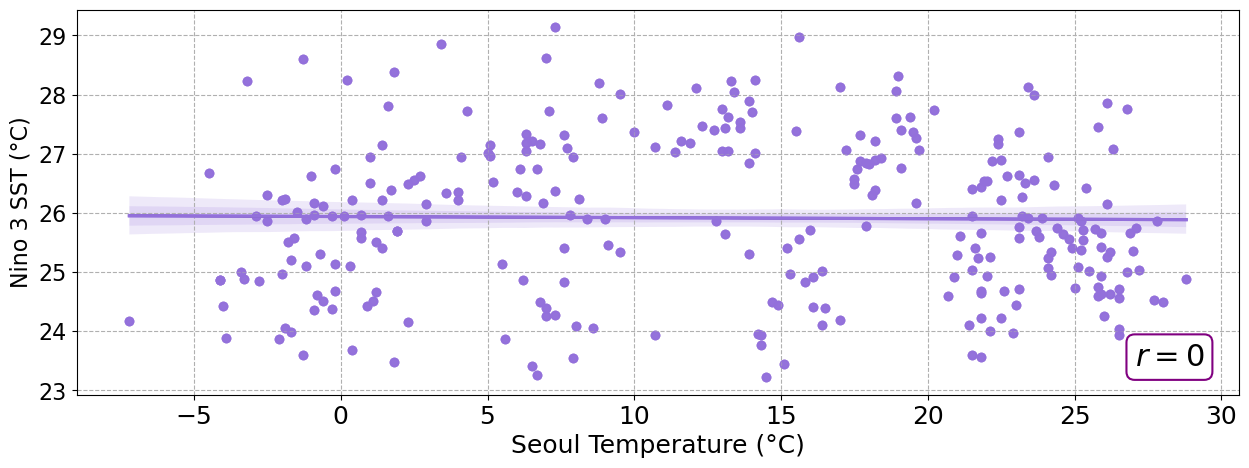}
    \includegraphics[width=8cm, height=5cm]{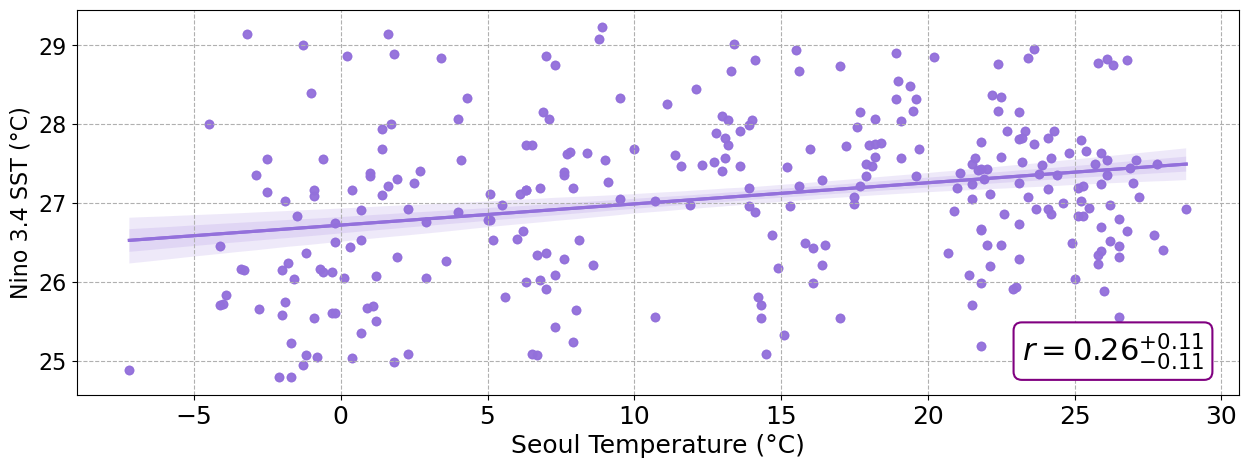}
    \includegraphics[width=8cm, height=5cm]{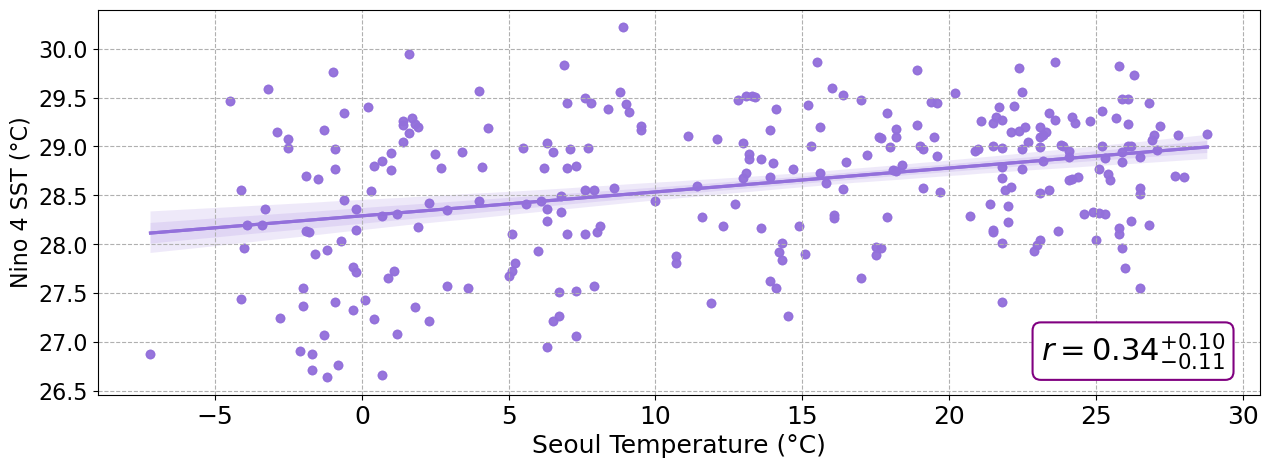}
    \caption{
	Correlation of average monthly temperatures in Seoul (1997-2020) with various Niño sea surface temperatures (SST). 
	The x-axis for all plots indicates Seoul's average monthly temperatures. 
	The y-axis of the graph, from top to bottom, represents Niño 1+2, Niño 3, Niño 3.4, and Niño 4. 
	If the significance level does note meet less than 0.05, the correlation coefficient is recorded as r=0. 
	The meaning of the r-values with the $\pm$ values are the same as Fig. 12.}
    \label{fig:SeoulTemp_NinoX_CORRs}
\end{figure}

Utilizing monthly average temperature data from 1997 to 2020 
for both Seoul and El Niño observation areas, Fig.\,\ref{fig:SeoulTemp_NinoX_CORRs} illustrates these correlations. 
Interestingly, moving from the eastern to the central Pacific, from the Niño 1+2 to the Niño 4 regions, 
we observe a transition in correlation 
from negative (Niño 1+2), to neutral (Niño 3), to positive (Niño 3.4 and Niño 4). 
Specifically, Niño 4 shows a stronger positive correlation than Niño 3.4. 
This pattern suggests that while Niño 3.4 is often highlighted for its importance in relation 
to Pacific phenomena like ENSO, both Niño 1+2 and Niño 4 regions 
also significantly influence Seoul's temperatures, challenging traditional perceptions.
Specifically, the Niño 1+2 region shows a pronounced negative correlation with Seoul’s temperature (-0.49), 
Niño 3.4 presents a weak positive correlation (0.26), and Niño 4 exhibits a more distinct positive correlation (0.34).
They highlight the complex and varied effects of the Pacific Ocean on temperatures in remote Seoul, 
and offer new insights into the global reach of El Niño-related temperature fluctuations.

\subsection{Correlation between SSTs in the South Sea and El Niño Monitoring Area}
\begin{figure}
    \centering
    \includegraphics[width=8cm, height=5cm]{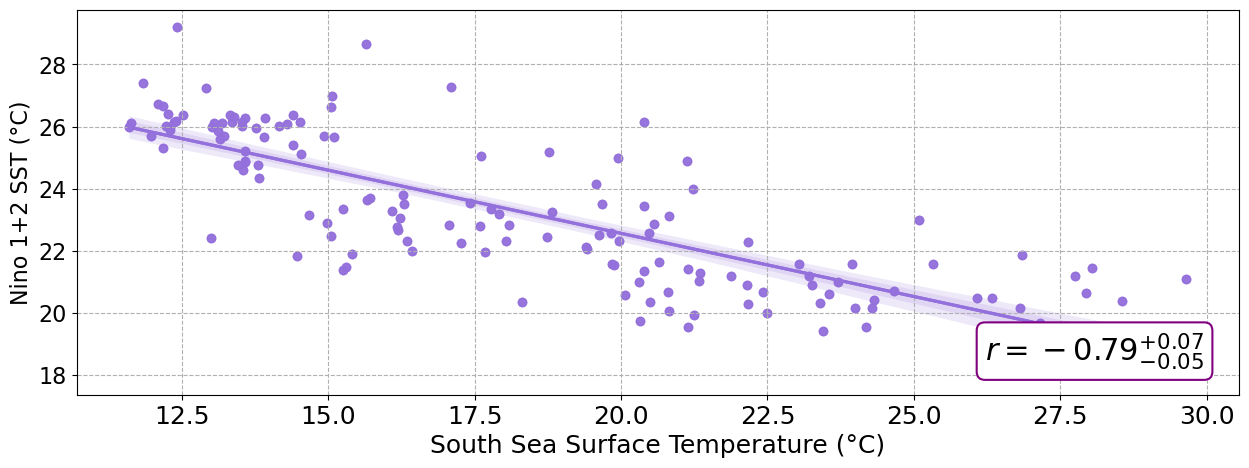}
    \includegraphics[width=8cm, height=5cm]{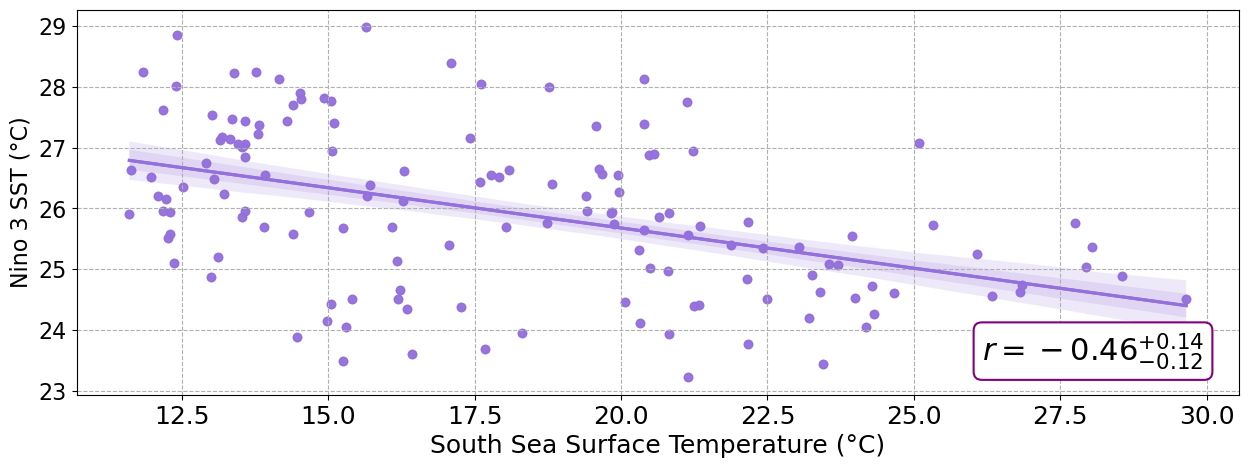}
    \includegraphics[width=8cm, height=5cm]{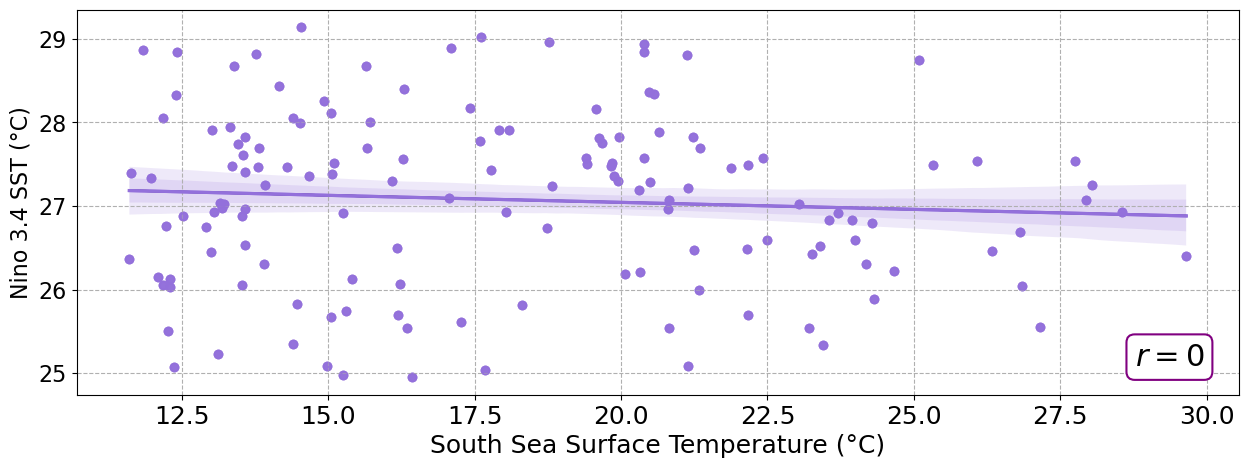}
    \includegraphics[width=8cm, height=5cm]{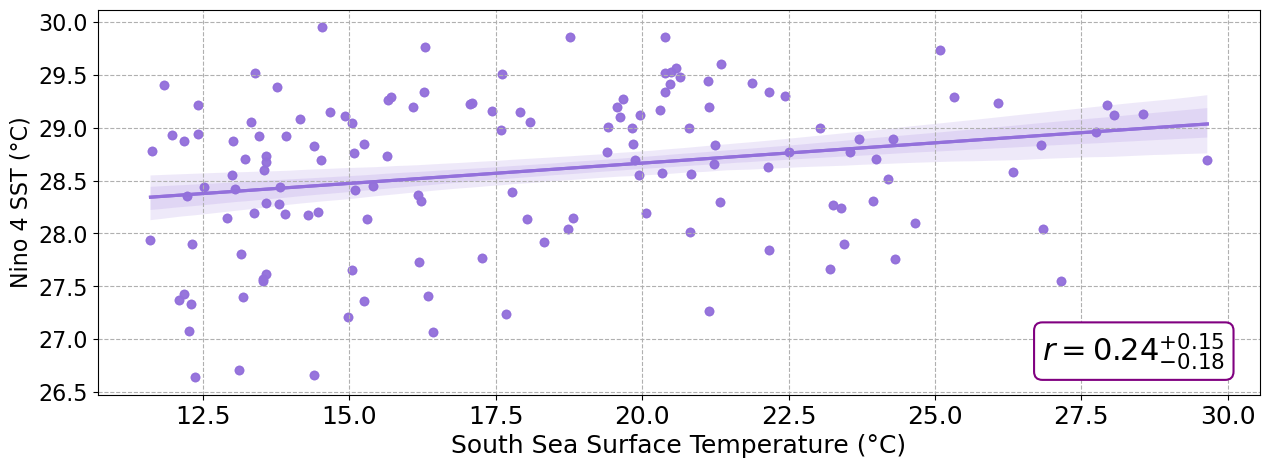}
     \caption{
	Three graphs detailing the relationship between South sea surface temperature (SST) 
	and various Niño regions from February 1997 to December 2020. Data points are sampled bi-monthly. 
    Top: y-axis showcases the SST of Niño 1+2.
    Middle: y-axis displays the SST of Niño 3.
    Bottom: y-axis illustrates the SST of Niño 4.
    SST denotes Sea Surface Temperature. 
    If the significance level is not below 0.05, the correlation coefficient is recorded as r=0. 
    The meaning of the r-values with the $\pm$ values are the same as Fig. 12.}
    \label{fig:SST_NinoX_CORRs}
\end{figure}

Fig.\,\ref{fig:SST_NinoX_CORRs} illustrates the correlation between South Sea SST and SSTs 
from El Niño monitoring areas Niño 1+2, Niño 3, and Niño 4. The x-axis for all graphs represents South Sea SST, 
while the y-axes show the SSTs for each Niño region. Each purple band illustrates the 1$\sigma$ and 2$\sigma$  uncertainties. 
The findings highlight a strong negative correlation (-0.79) with Niño 1+2, a slight negative correlation (-0.46) 
with Niño 3, and a weak positive correlation (0.24) with Niño 4, indicating a particularly pronounced negative correlation 
in the Niño 1+2 region compared to Niño 3 and Niño 4.

Interestingly, when previously analyzing Seoul temperatures, the correlation sign changed 
from negative to positive near the Niño 3 region, but this time the sign changes near the Niño 3.4 region.
This suggests that the Niño 1+2 (strong negative correlation) and Niño 4 (positive correlation) regions are 
the only El Niño monitoring areas consistently showing the same correlation pattern with both distant Seoul and the South Sea simultaneously. 
This consistency indicates that changes in the South Sea SST are influenced by similar oceanic processes 
affecting the Niño 1+2 and Niño 4 regions, thereby impacting distant locations like Seoul.

\subsection{Correlations of Niño 1+2 and Niño 4 SST with respect to Niño 3.4 SST}
\begin{figure}
    \centering
    \includegraphics[width=8cm, height=5cm]{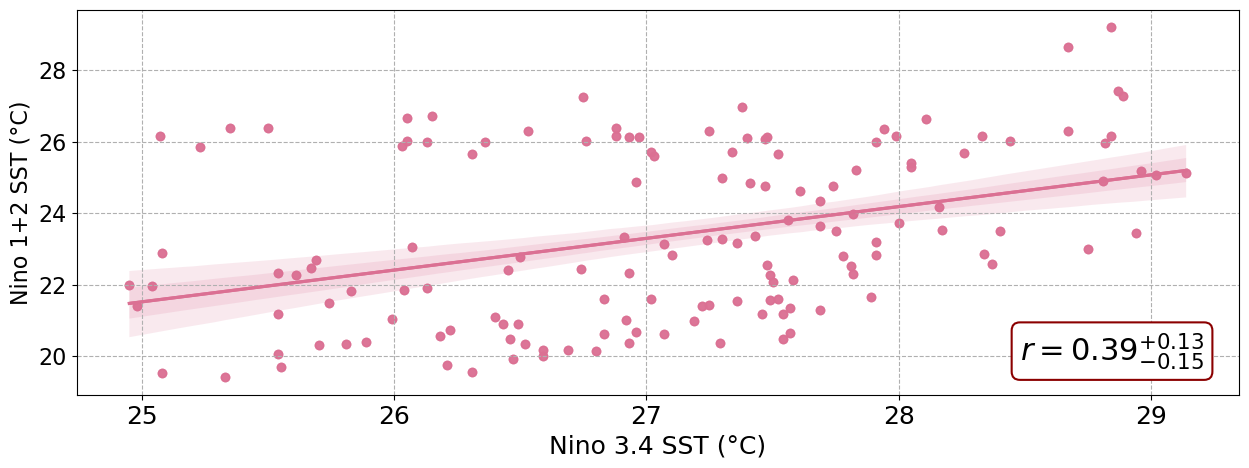}
    \includegraphics[width=8cm, height=5cm]{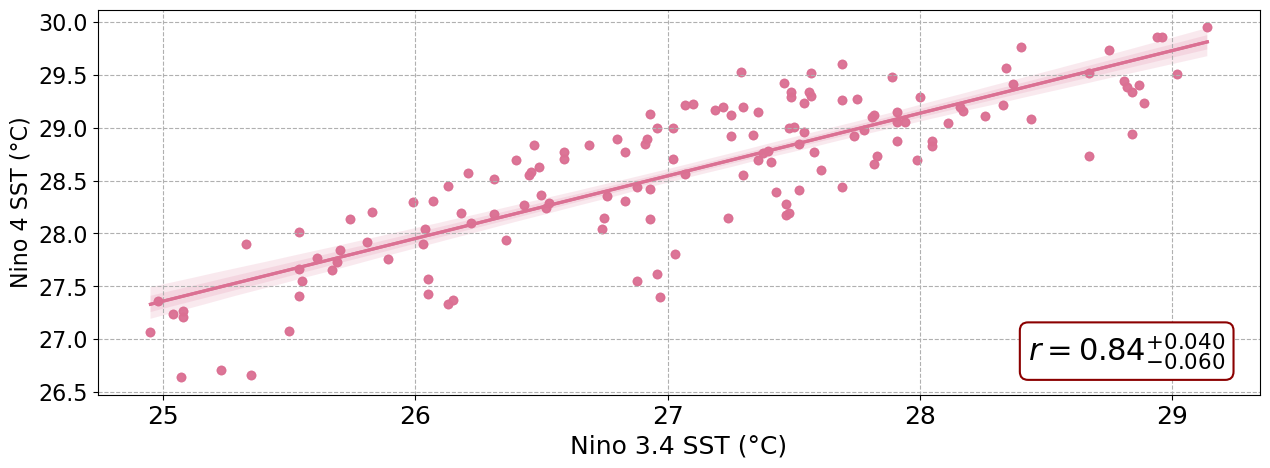}
    \caption{
   	Scatter plots displaying correlations between sea surface temperatures 
   	of Niño 3.4 (x-axis) with Niño 1+2 (left y-axis) and Niño 4 (right y-axis). 
   	Data spans from February 1997 to December 2020 in two-month intervals. 
   	The meaning of the r-values with the $\pm$ values are the same as Fig. 12.}
    \label{fig:NinoX_Nino3.4_CORRs}
\end{figure}

Our study focuses on SSTs in the Niño 1+2 and Niño 4 regions, 
because they can have positive or negative correlations with both Seoul temperature and South Sea SST at the same time. 
This highlights the significant link between these regions despite their geographical separation. 
In this context, we examine the correlation between SSTs in these areas and the SST in the Niño 3.4 region. 
Our results, illustrated in \ref{fig:NinoX_Nino3.4_CORRs}, 
feature scatterplots with Niño 3.4 SST on the x-axis and SSTs from Niño 1+2 (top panel) and Niño 4 (bottom panel) on the y-axis. 
Each plot includes a red line indicating a positive linear correlation, 
with a surrounding band that represents the 1$\sigma$ and 2$\sigma$ uncertainty levels. 
Given the Niño 3.4 region's role as a transitional zone between Niño 3 and Niño 4, 
it is expected to have a stronger correlation and lower uncertainty with Niño 4 than with Niño 1+2. 
Indeed, the correlation with Niño 4 is found to be +0.84 (with an uncertainty range of +0.040 to -0.060), 
significantly stronger and with less uncertainty than the +0.39 correlation (with an uncertainty range of +0.13 to -0.15) 
with the Niño 1+2 region, as shown in Fig\,\ref{fig:NinoX_Nino3.4_CORRs}. 
These correlations are crucial for developing models to predict the SEI in following sections.
\\ \\
These notable correlation results are a key consideration in developing a model for determining the SEI, 
as introduced in the following section. 
For a more detailed correlation analysis and methodologies, see the Appendices of this paper.

\section{Model for Super El Niño prediction}
To create a model to predict SE, we assume that the effects of SE could extend 
beyond the Pacific Ocean and the Pacific Rim and have a significant impact on Korea and Japan in the Northern Hemisphere. 
More specifically, when a SE occurs:
\begin{enumerate}
\item It affects ocean currents from western South America near the equator to the eastern and central Pacific and eastern Asia.
\item This effect extends to Korea and Japan in the Northern Hemisphere along the Kuroshio Current, 
      which originates in the North Equatorial Current.
\end{enumerate}
In addition to these assumptions, the following factors are crucial for building an accurate SE prediction model:
\begin{enumerate}
\item Correlations with South Sea SST and Seoul Temperatures.
They indicate the far-reaching impacts of SE events on the Northern Hemisphere. 
These correlations are integral to our model.
\item Roles of Niño 1+2 and 4 regions. 
They have shown significant correlations with both South Sea SST and Seoul temperatures. 
The strong negative correlation with Niño 1+2 and the positive correlation with Niño 4 are particularly important. 
These regions are weighted heavily in the model due to their meaningful impact on both the South Sea and Seoul.
\item Historical Data Analysis. 
Utilizing historical data from previous SE events (1982-83, 1997-98, and 2015-16) is essential for validating the model. 
The patterns and correlations observed during this period provide 
valuable information about the conditions under which SE events occur.
\end{enumerate}
Based on the correlation results and factors discussed in Sec.\,\ref{sec.:CA}, 
the following subsections present a comprehensive model that can predict SE events.

\subsection{Determining the Correlation Coefficients of the Model}
The Niño 3.4 region has been identified as a crucial area for tracking El Niño occurrences. 
By assuming that reactions in this region extend to the Niño 1+2 and Niño 4 regions, 
and acknowledging that only Niño 1+2 and Niño 4 exhibit consistent correlations with distant areas of the Northern Hemisphere, 
we can deduce the significant model coefficients related to Seoul and the South Sea SST.

Our analysis begins with examining the correlation between the Niño 1+2 (or Niño 4) region and the Niño 3.4 region. 
The model coefficient is determined by the ratio of each correlation coefficient to the sum of the correlation coefficients. 
For instance, the coefficient from Niño 3.4 to Niño 1+2 is derived by dividing its correlation coefficient (0.39) 
by the sum of the correlation coefficients (0.39 + 0.84), resulting in a value of 0.32. 
Similarly, the coefficient for Niño 4 (0.84) becomes 0.68.

Next, the coefficient between Seoul (or South Sea) temperature and Niño 1+2 SST is calculated. 
For Seoul temperature through Niño 1+2, the correlation coefficient (0.49) is divided 
by the sum of the correlation coefficients (0.49 + 0.79), resulting in a value of 0.38. 
For South Sea temperature through Niño 1+2, the coefficient is 0.62.

Starting from Niño 3.4, the model coefficient for Seoul (or South Sea) temperature 
through Niño 1+2 is computed by multiplying the coefficient from Niño 3.4 to Niño 1+2 (0.32) 
by the coefficient from Niño 1+2 to Seoul (or South Sea) temperature (0.38 for Seoul and 0.62 for South Sea). 
This results in values of 0.12 for Seoul and 0.20 for South Sea. Similarly, from Niño 3.4 SST through Niño 4, 
the model coefficient for Seoul (or South Sea) temperature is calculated 
by multiplying the coefficient from Niño 3.4 to Niño 4 (0.68) by the ratio of the correlation coefficients 
between Niño 4 and Seoul (or South Sea) temperature (0.36 for Seoul and 0.24 for South Sea) 
divided by their sum. This results in values of 0.41 for Seoul and 0.27 for South Sea.

\begin{figure*}
    \centering
    \includegraphics[width=12cm, height=6cm]{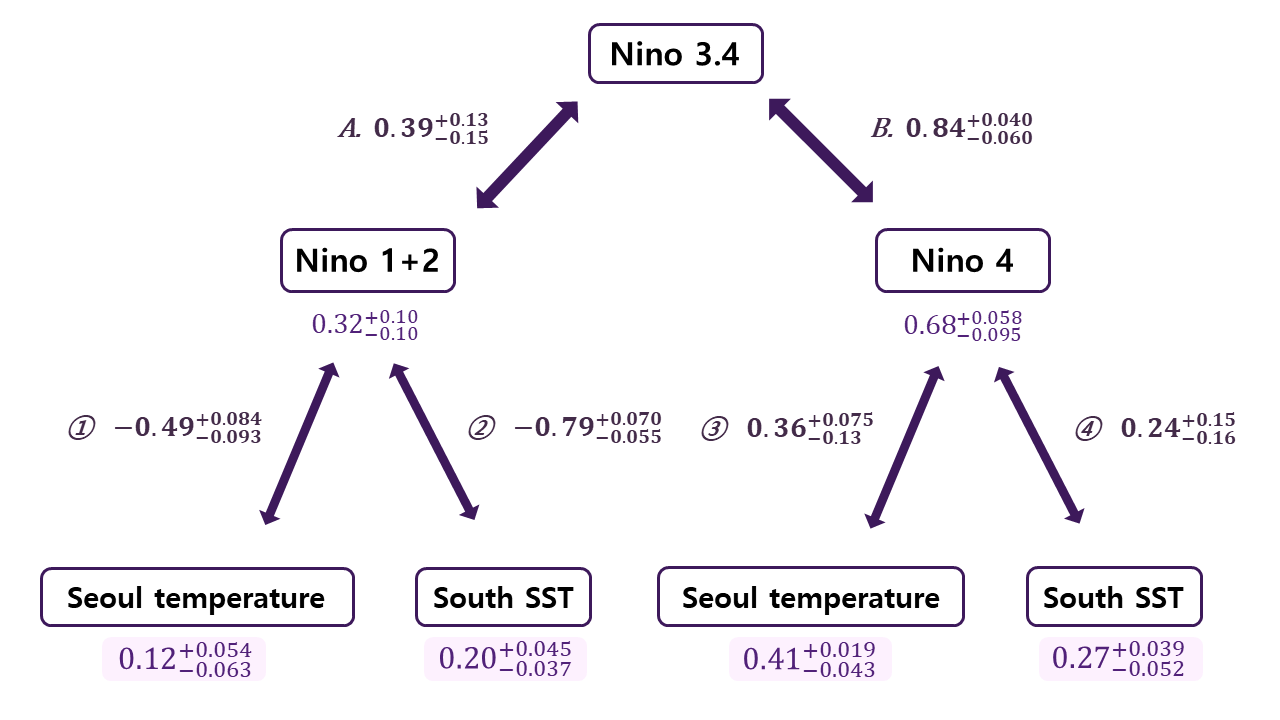}
    \caption{
This diagram aggregates the correlation coefficients from Fig.\,\ref{fig:SSHs_CORRs}, \,\ref{fig:SST_SSH_CORRs},
 \,\ref{fig:SeoulTemp_SST_CORR}, \,\ref{fig:SeoulTemp_NinoX_CORRs}, \,\ref{fig:SST_NinoX_CORRs} and \,\ref{fig:NinoX_Nino3.4_CORRs}
to derive the model coefficients for Eq. (\ref{eq:SEI}). 
Given the parallel trends in Niño 3.4 (the benchmark for El Niño assessment in Korea), Niño 1+2, and Niño 4, 
the correlation coefficients for Seoul temperature and South SST are portrayed using their respective correlation coefficients. 
Symbols $\pm$ within A, B, 
\raisebox{.5pt}{\textcircled{\raisebox{-.9pt} {1}}}
\raisebox{.5pt}{\textcircled{\raisebox{-.9pt} {2}}}
\raisebox{.5pt}{\textcircled{\raisebox{-.9pt} {3}}}, and
\raisebox{.5pt}{\textcircled{\raisebox{-.9pt} {4}}} indicate the 95$\%$ confidence interval for the Pearson correlation coefficient.}
    \label{fig:15_all_coefficients}
\end{figure*}

Figure \ref{fig:15_all_coefficients} summarizes the corresponding model coefficients 
for Seoul and South Sea temperatures from Niño 3.4 SST. In this figure, 
$A(B)$ adjacent to the arrow signifies the linear correlation coefficient and $2\sigma$ uncertainty between the Niño 3.4 and Niño 1+2(4) regions, 
while \raisebox{.5pt}{\textcircled{\raisebox{-.9pt} {1}}}(\raisebox{.5pt}{\textcircled{\raisebox{-.9pt} {3}}}) and 
\raisebox{.5pt}{\textcircled{\raisebox{-.9pt} {2}}}(\raisebox{.5pt}{\textcircled{\raisebox{-.9pt} {4}}}) 
denote the  Niño 1+2 (Niño 4) SST, and the Seoul temperature and South Sea SST, respectively. 
The purple numerals and uncertainties beneath each box reflect model coefficients derived from these linear correlations.
These four model coefficients are essential components used in the model for predicting SE events.

\subsection{Introduction to Model}
Figure\,\ref{fig:16_all_correlations} summarizes the correlation coefficients 
between Seoul temperature, South Sea SST, and each Niño region as indicated in the boxes. 
The numbers above (or below) each Niño region denote the correlation coefficients with Seoul (or South Sea SST). 
The 'X' mark indicates that no significant correlation appears between the two regions.
We can easily see that the Niño 3 has no significant correlation with Seoul temperature, and the Niño 3.4 also has no correlation with South Sea SST (See Fig.\,\ref{fig:SeoulTemp_NinoX_CORRs} and \,\ref{fig:SST_NinoX_CORRs}), 
while the Niño 1+2 region shows a negative correlation with both regions, 
and the Niño 4 region shows a positive correlation. 
Therefore, considering the remote correlations, 
it is understandable why only the Niño 1+2 and Niño 4 regions were selected for the present model predicting SE.

Using the four correlation coefficients mentioned above (summarized in Fig.\,\ref{fig:16_all_correlations}) and their corresponding model variables, 
we can construct a model to predict SE as shown below:
\begin{figure}
    \centering
    \includegraphics[width=8cm, height=5cm]{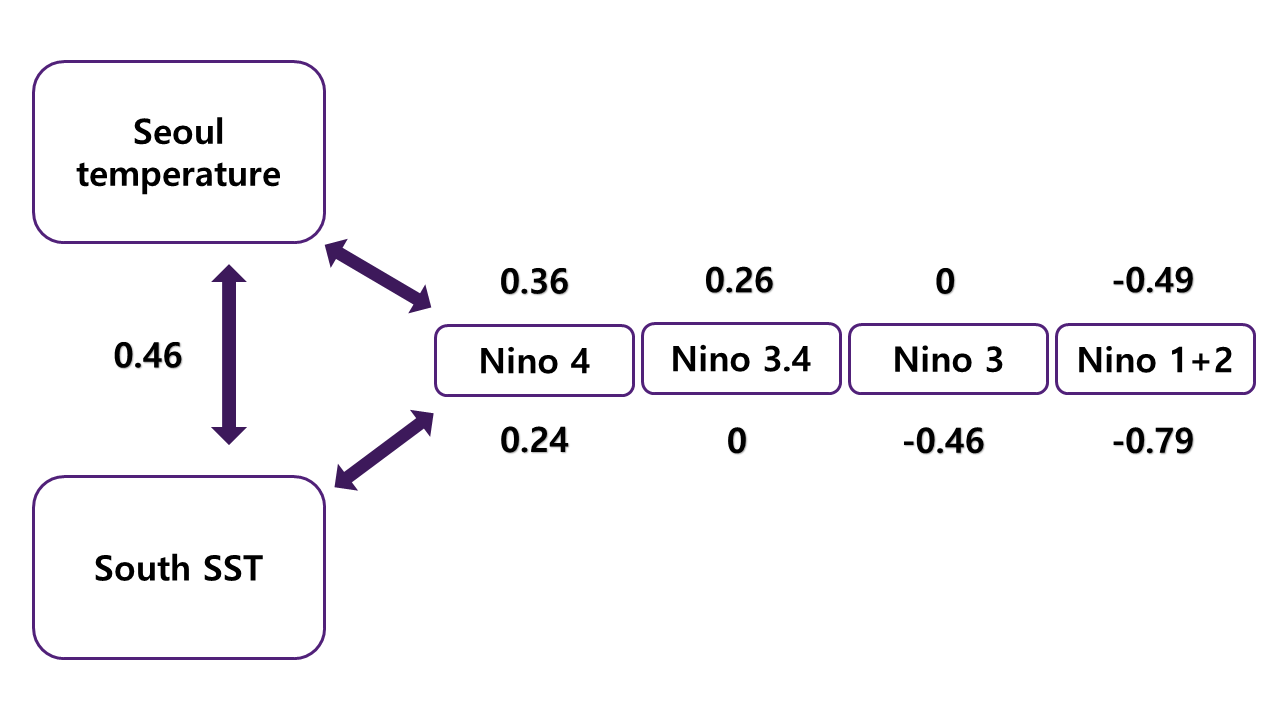}
    \caption{Correlation coefficients for Seoul temperature, South sea surface temperature, 
    	and El Niño monitoring areas (Niño 1+2, Niño 3, Niño 3.4, and Niño 4). 
    	SST denotes Sea Surface Temperature. 
        The two 'zero's comes from the correlation coefficients r=0 in Fig.\,\ref{fig:SeoulTemp_NinoX_CORRs} and \,\ref{fig:SST_NinoX_CORRs}.
        If the significance level does not fall below 0.05, the correlation coefficient is designated as '0'. 
    }
    \label{fig:16_all_correlations}
\end{figure}

\begin{equation}
\centering
E = 0.41\,X^{\rm Nino 4}_{\rm Seoul} + 0.27\,Y^{\rm Nino 4}_{\rm South} 
+ 0.12\,Z^{\rm Nino 1+2}_{\rm Seoul} + 0.20\,W^{\rm Nino 1+2}_{\rm South}~,
\label{eq:SEI}
\end{equation}
where the model variable X(Z) denotes the average winter temperature in Seoul for the months of December to February, 
two years (four months) prior to the occurrence of El Niño. Y(W) represents the South Sea SST one month (six months) 
before El Niño manifests. The terms X and Y are associated with Niño 4, while Z and W pertain to Niño 1+2.
Here we explain why we choose the temperature for the fixed period. Fig.\,\ref{fig:Seoul_Temp_Nino_deviation_normalized} in Appendix A presents several key comparisons.
The first graph depicts Niño 1+2 SST and Seoul temperature from 1997 to 2020. The peaks of these two graphs consistently show a phase difference of approximately 4 months, which can be described as multiples of 4 months .
The third graph illustrates Niño 1+2 SST and South SST over the same period. These graphs exhibit a similar pattern with intervals of approximately 6 months, described as multiples of 6 months.
The fourth graph indicates that Niño 4 SST and South SST share a similar pattern with intervals of roughly 1 month, described as multiples of 1 month.
The second graph suggests that lower winter temperatures in Seoul two years prior (2 years prior winter) can influence the occurrence of subsequent Super El Niño events.
Consequently, the minimum values of each cycle were utilized to determine the model's X, Y, Z, and W parameters. For more detailed information, please refer to Appendix A.

\begin{figure*}
	\centering
        \includegraphics[width=18cm, height=10cm]{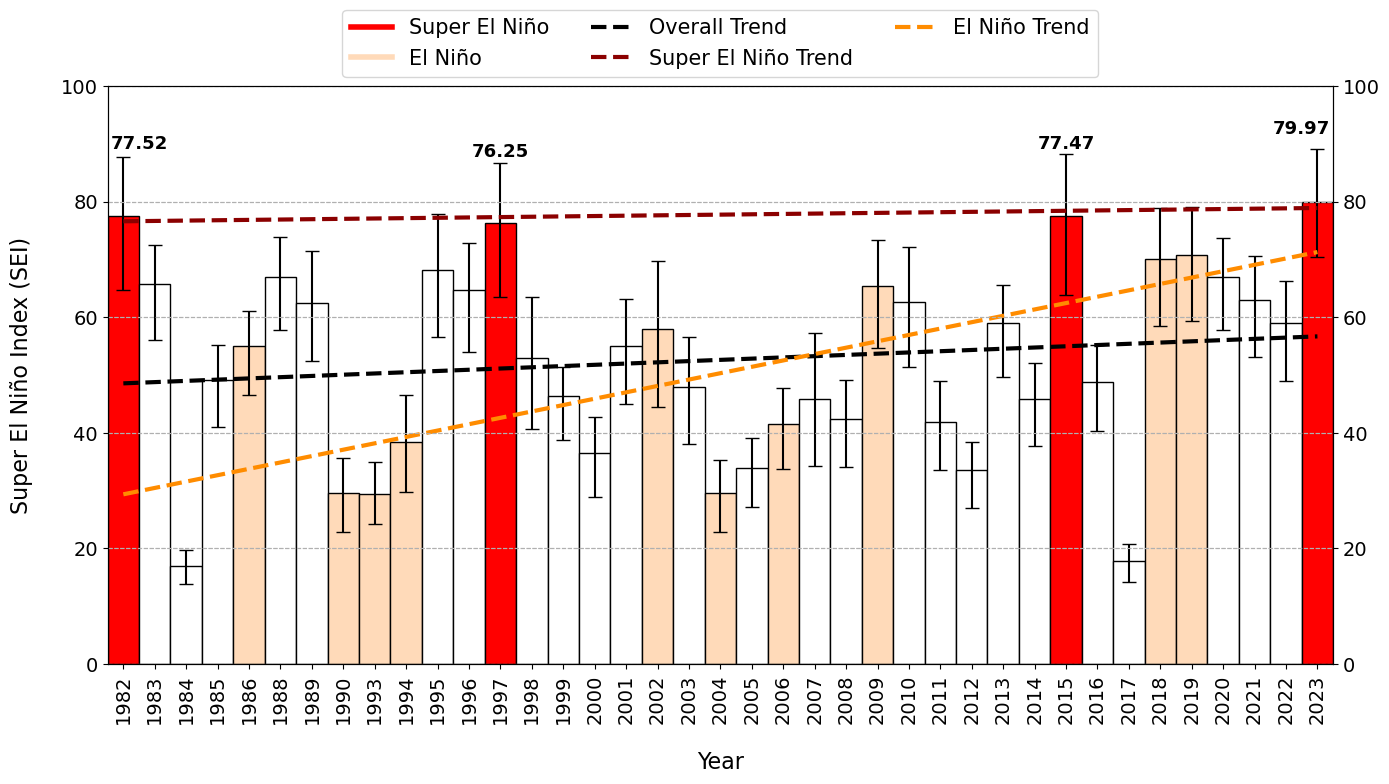}
	\caption{
Super El Niño Index (SEI) values over the years. Super El Niño events are marked with red boxes, while regular El Niño events are indicated by orange boxes. White boxes represent SEI values for periods that are neither Super El Niño nor El Niño.
The dashed black line shows the overall trend through linear fitting of the data, indicating an increasing SEI since 1980, suggesting a rise in teleconnection correlations. The dark red dashed line highlights the trend associated with Super El Niño events, and the orange dashed line pertains solely to regular El Niño events. The black error bars on each bar represent the 2$\sigma$ uncertainty.
The increasing trend in El Niño suggests that by around 2030, the trend may converge to the rising trend of Super El Niño, indicating a higher probability of regular El Niño events transitioning into Super El Niño events. Furthermore, the trend line associated with Super El Niño shows stability at an SEI value around 80, suggesting the reliability of these SEI values.
The SEI values for periods that are neither Super El Niño nor regular El Niño are estimated by repeating the SEI values approximately 12-13 months after the end of the previous event. This 12-13 month interval is derived from the average duration of past El Niño and Super El Niño events. For more detailed information, please refer to the supplementary Table.}
	\label{fig:SEI2}
\end{figure*}

The model's value (E) is then scaled by multiplying by 100 to create the SEI, 
which ranges from 0 to 100. Values closer to 100 indicate greater remote correlations, 
reflecting a characteristic of SE. The values obtained by this model 
for previous (super) El Niño events are shown in Fig.\,\ref{fig:SEI2}.

Figure \ref{fig:SEI2} presents the SEI values over the years from 1982 to 2023.
For this index, the known (super) El Niño period was used for the years with (super) El Niño, 
and the period after the (super) El Niño period was considered for the years without (super) El Niño.
SE events are distinguished by red rectangular boxes, 
while regular El Niño events are marked with orange rectangular boxes. 
The black line across the graph depicts the overall trend based on the linear fitting of these values, 
indicating an upward trend in SEI since 1982. This trend suggests a steadily increasing pattern of remote correlation, 
possibly reflecting the influence of global warming on the frequency and intensity of SE events.

Notably, the median SEI values for SE events are all above 75, demonstrating the model's accuracy in identifying these events. 
In contrast, many regular El Niño events have much lower SEI values, highlighting the model's distinction 
between Super El Niño and regular El Niño events. 
This distinction shows the model's capability in effectively predicting SE occurrences.

Interestingly, the model's coefficients were determined using only data up to 2022, without considering the 2023 SE data. 
Nevertheless, our model predicted a very strong El Niño in 2023, with the value reaching $80^{+9}_{-10}$. 
The next section will briefly discuss the validation of this prediction.

Another meaningful observation is that the SEI values for regular El Niño events have approached around 70 since 2015. 
While the exact cause of this trend is unclear, it could imply that the increasing global temperatures are 
making conditions more conducive to the occurrence of strong El Niño events. 
This increasing trend of SEI highlights the need for continuous monitoring and adaptation strategies 
to address the potential rise in frequency and intensity of SE events in the near future.

\section{Validation of 2023 Super El Niño Predictions}\label{se6}
\begin{figure*}
	\centering
	\includegraphics[]{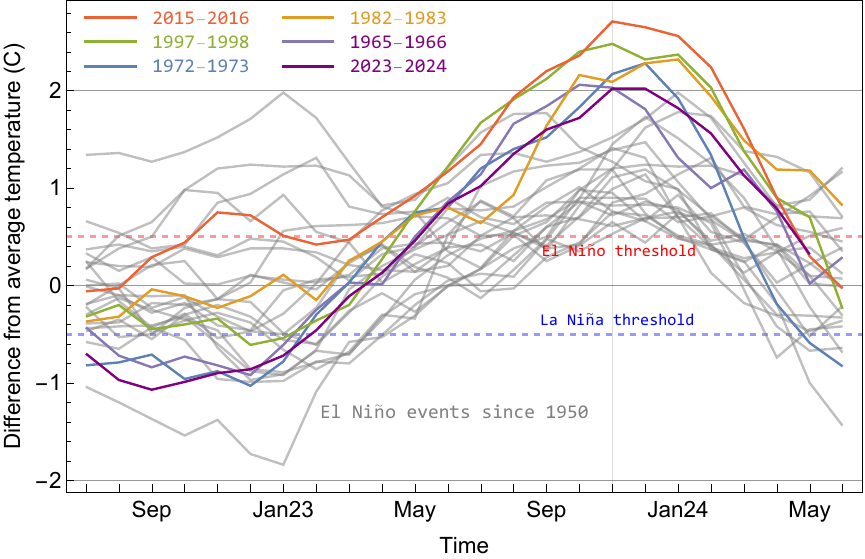}
	\caption{Temperature deviations from the 30-year average for all El Niño events since 1950, with a focus on the prediction validation for the 2023-24 Super El Niño event using SST data from July preceding the El Niño through January of the following year. It highlights the temperature differences beginning to escalate from the average in El Niño years, peaking in November, then declining towards a neutral range (-0.5 to +0.5) by June post-El Niño. It also shows five Super El Niño events in varied colors against the standard El Niño events in gray. Notably, by November, Super El Niño events demonstrate temperature anomalies surpassing 2 degrees, distinctly separating them from other events typically at or below 1.5 degrees. 
	The deviation shown by the purple line from July 2021 to January 2024 conclusively indicates the occurrence of a Super El Niño in the 2023 season, marking it as the fifth or sixth event of its magnitude in historical records. This confirms the model's accuracy in predicting Super El Niño events, as evidenced by the high SEI value for 2023, validating the model's robustness for such predictions.}
	\label{fig:DAT_Nino3.4}
\end{figure*}

This section confirms the validation of the model's prediction 
for a SE event during the 2023-24 period. 
To achieve this, SST data from July of the year preceding the El Niño occurrence
through January of the subsequent year is used to estimate deviations from the 30-year average temperatures. 
The deviations for all El Niño events since 1950 are illustrated in Fig.\,\ref{fig:DAT_Nino3.4} 
(note that ONI, which uses a three-month average, is not used for the purpose of indicating the exact period 
satisfying the specific condition, namely +2.0).

Typically, in El Niño years, the temperature difference from the average begins to rise, peaking around November, 
before starting to decrease towards a neutral range (-0.5 to +0.5) by June of the year following El Niño. 
Importantly, among these, there were five SE events, which are highlighted in different colors 
to distinguish them from the standard El Niño events (depicted in gray). 
By November, Super El Niño events and normal El Niño events are clearly differentiated based on a temperature difference of 2 degrees. 
To emphasize this point, the gray vertical line in Figure 21 is set to November.

Figure \ref{fig:DAT_Nino3.4} illustrates the sequence of SE events 
based on the maximum temperatures in November during the SE period: 
2015-16 (red), 1997-98 (green), 1972-73 (blue), 1982-83 (orange), 1965-66 (light purple), and 2023-24 (purple). 
The difference in average temperatures from July 2021 to January 2024, represented by the purple line, 
clearly confirms that an SE event occurred in 2023, ranking it as the fifth most intense SE since the 1950s. 
As this value has recently returned to the neutral range, it indicates that the 2023 SE has begun its decline phase. 
Additionally, based on historical patterns, a La Niña event is expected to follow soon.

This validation demonstrates the robustness of the model for predicting SE events, 
highlighting its potential utility in future climate forecasting efforts.

\section{Conclusion}
In this study, we developed a robust predictive model for SE events by integrating both local and global climatic data. 
Our model introduces the SEI, which effectively captures the occurrence and intensity of SE events. 
This model provides significant improvements over traditional equatorial-centric forecasting methods, demonstrating high accuracy and reliability.

Our research made several key findings and contributions. 
Firstly, we introduced a novel predictive model and SEI. 
Our model incorporates data from diverse sources, including SSTs from traditional El Niño monitoring areas, 
SSTs and sea level measurements around the Korean Peninsula, and temperature records from Seoul. 
The SEI, with a threshold value of 80, accurately distinguishes between regular El Niño and SE events. 
Historical data validation confirms the model's feasibility to identify major SE events, including those from 1982-83, 1997-98, and 2015-16.

Secondly, we identified significant remote correlations. 
The analysis highlights substantial correlations between SSTs in the Niño1+2 and Niño4 regions 
and climatic conditions in distant regions like Seoul and the South Sea. 
These correlations emphasize the far-reaching impacts of SE events, extending beyond the equatorial Pacific.

Thirdly, our findings indicate an increasing frequency and intensity of SE events. 
The consistent increase of SEI values since 1982 suggests that conditions favorable to SE events are becoming more frequent. 
This trend is likely driven by global warming, indicating a need for global preparedness and strategic planning to mitigate the impacts of future SE events.

Fourthly, the validation of our model and its future predictions underscore its accuracy and reliability. 
The model's prediction of the 2023 SE event, confirmed by recent observations, 
ranks this event as the fifth most intense since the 1950s, reflecting the model's capability in capturing the nuances of SE occurrences.

Lastly, our study has significant implications for climate strategy. 
The increasing trend in SEI values and the narrowing gap between regular El Niño and SE events point to 
the potential for more frequent and intense SE events in the future. 
This necessitates the development of adaptive strategies to handle the adverse impacts on human health, infrastructure, and economies.

In conclusion, the SEI model represents a significant advancement in the prediction of Super El Niño events, 
offering valuable insights into the dynamics of global climatic changes leading to remote and global correlations. 
Accurate predictions of SE events can facilitate better preparation and mitigation of the associated extreme weather conditions. 
Our research emphasizes the urgent need for global adaptation strategies 
to address the increasing frequency and intensity of SE events driven by climate change.

Future research should focus on refining the model 
to further enhance its predictive accuracy and exploring the complex interactions between global warming and SE events. 
Developing more effective mitigation and adaptation strategies will be crucial 
in minimizing the adverse impacts of these increasingly common climatic phenomena. 
The continued study of SE events and their broader implications will contribute to 
a deeper understanding of global climate dynamics and help policy-making for climate resilience.

\section*{Acknowledgements}
This research was supported by a grant from the National Research Foundation of Korea (Grant No. NRF-2020R1A2C3006177, 
Grant No. NRF-2021R1A6A1A03043957, and 2018R1D1A1B07051126).

\appendix
\section*{Appendix}
In Appendix A, we discuss critical aspects among the data sets directly related to our model. Appendices B, C, and D provide detailed explanations and supplementary analyses that support our primary research findings. Each section contains comprehensive information on the methodologies, data sets, and models employed in our study.

\section{Time-series graphs from 1997 to 2020 comparing normalized temperature deviations.}
\begin{figure*}
	\centering
	\includegraphics[width=14cm, height=10cm]{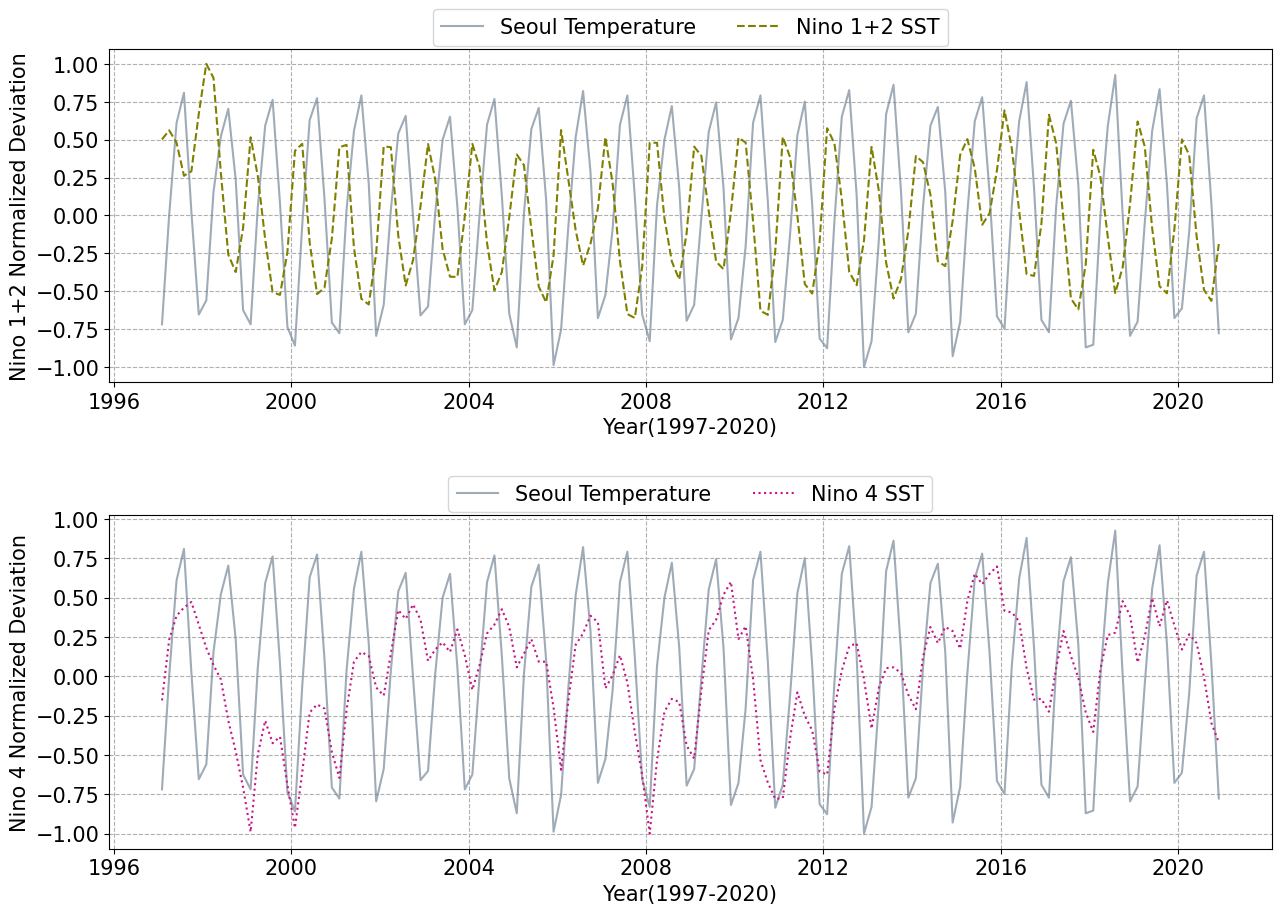}
	\includegraphics[width=14cm, height=10cm]{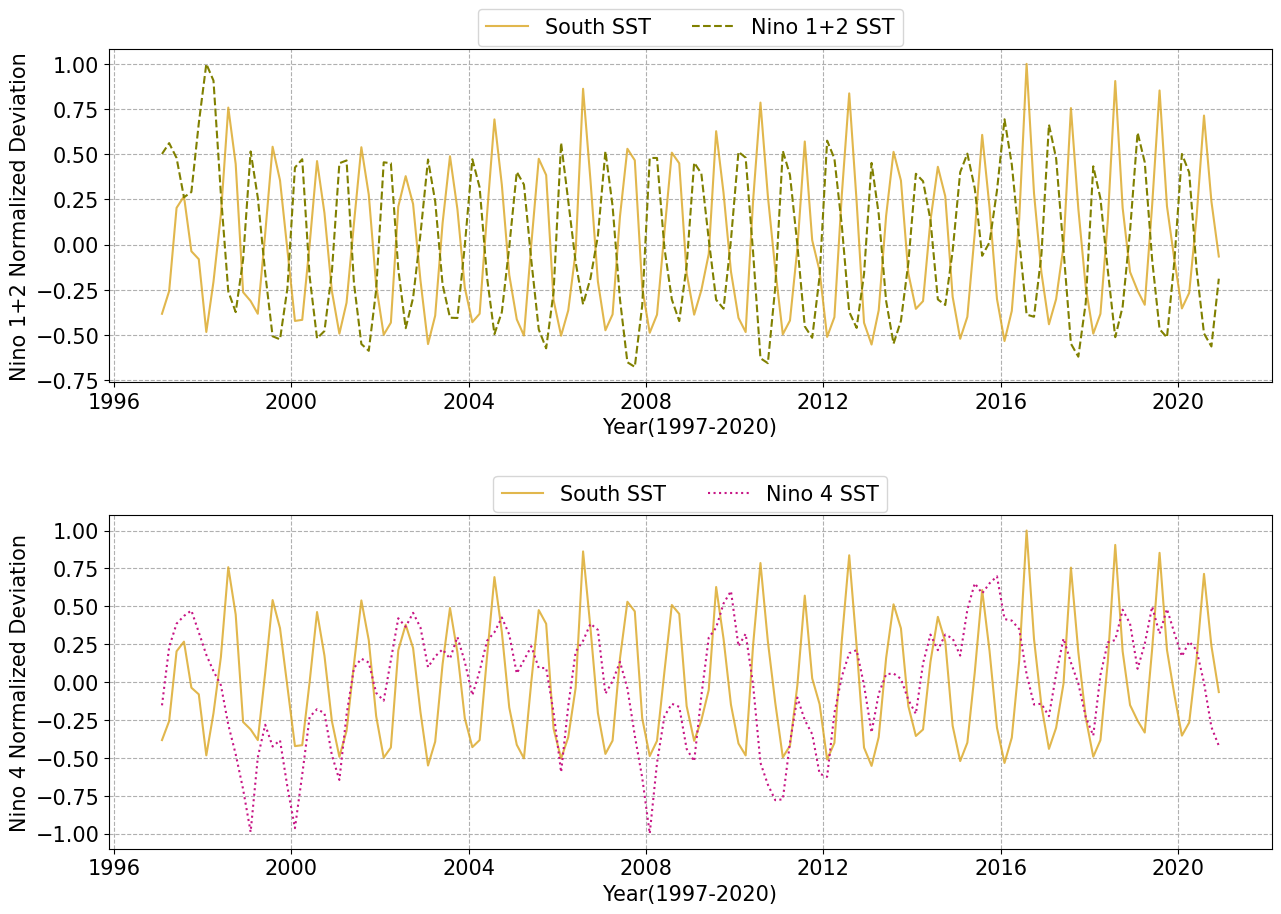}
	\caption{Time-series graphs from 1997 to 2020 comparing normalized temperature deviations of Niño 1+2 and Niño 4 SST to Seoul temperature and South SST. In the first two graphs, the light blue solid line depicts normalized deviations from the average Seoul temperature. In the latter two, the yellow line represents normalized deviations of the South sea surface temperature. The green dotted lines and pink dotted lines, across all graphs, denote normalized deviations of the sea surface temperature for Niño 1+2 and Niño 4, respectively. All values were normalized by subtracting the average and dividing by the respective maximum. The prominent regular deviation of Niño 1+2 to Seoul temperature and South SST facilitated a clear visualization of temperature tendencies. Their correlations in each panel are presentd in Fig.\,\ref{fig:SeoulTemp_NinoX_CORRs} and \,\ref{fig:SST_NinoX_CORRs}.}
	\label{fig:Seoul_Temp_Nino_deviation_normalized}
\end{figure*}

Fig.\,\ref{fig:Seoul_Temp_Nino_deviation_normalized} illustrates the relationships by mean-centering and normalizing the data, which highlights these correlations more clearly. Notably, it underscores the inverse correlation between Niño 1+2 and temperatures in Seoul and the South Sea. Additionally, it reveals a significant but less pronounced relationship between Niño 4 and these temperatures. This finding suggests that sea surface temperature (SST) data from Seoul and the South Sea may be useful in predicting El Niño events. The distinct behaviors of Niño 1+2 and Niño 4 provide valuable insights into the complex interplay of global climate dynamics and their impact on temperature patterns in Korea.

\section{Analysis of Time Series Data: Evaluating Stationarity and Implementing Forecasting Models}
Time series analysis is pivotal for visualizing and predicting future trends by drawing on historical data. While the ARIMA model excels at forecasting the near future, its predictive power wanes for long-term predictions\,\cite{M. Wang_2022_ARIMA}. Consequently, this study incorporated a Bayesian approach for time series modeling, with the outcomes of Bayesian linear regression using the dynesty analysis detailed in Appendix C.

Tables \ref{tab:Seoul temp using ARIMA BSTS}, \ref{tab:EWS SST ARIMA BSTS}, \ref{tab:EWS SSH LR BLR}, \ref{tab:EWS actual and predic SSH}, \ref{tab:ARIMA_Seoul}, \ref{tab:BSTS_Seoul}, and \ref{tab:BSTS_EWS_SST} collectively summarize our findings:

\begin{table}[hbt!]
	\begin{tabular}{m{1.5cm} m{3.4cm} m{3.4cm}}
		\hline
		\vspace{0.5ex}
		Model        & ARIMA      & BSTS  \\ \hline \hline
		\vspace{0.5ex}
		MAPE        & 7.98$\%$ & 3.82$\%$  \\  \hline
		\vspace{0.5ex}
		Linear        & \multirow{2}{*}{y = $4.29*10^{-4}$x + 13.0} & \multirow{2}{*}{y = $1.06*10^{-2}$x + 13.03} \\
		Regression &             & \\     \hline
		\vspace{0.5ex}
	\end{tabular}
	\caption{
	Analysis of temperature trends in Seoul. The Mean Absolute Percentage Error (MAPE) and linear regression are utilized to examine the data. Additionally, forecasts using the Bayesian Structural Time Series (BSTS) model predict a more significant rise in temperatures for Seoul.}
	\label{tab:Seoul temp using ARIMA BSTS}
\end{table}

The tables \ref{tab:Seoul temp using ARIMA BSTS} provides a detailed comparison of the Mean Absolute Percentage Error (MAPE) between the actual and fitted values, as well as the linear regression equations for different models. The results indicate that the MAPE for the Bayesian Structural Time Series (BSTS) model is roughly half of that for the ARIMA model, highlighting the superior accuracy of the BSTS approach. Additionally, the slope of the ARIMA model's linear regression equation is about 100 times smaller than that of the BSTS model. This significant difference suggests that the BSTS model not only offers more precise predictions but also forecasts a substantially greater increase in Seoul's temperature compared to the ARIMA model. Therefore, the BSTS model is more effective for anticipating future temperature trends in Seoul.

\begin{table*}[]
	\setlength{\tabcolsep}{1.5pt}
	\renewcommand{\arraystretch}{1}
	\begin{tabular}{m{2cm} m{4cm} m{5cm} m{5cm}}
		\hline
		\vspace{0.5ex}
		SST                  &Model                   &ARIMA                   &BSTS                   \\ \hline \hline
		\vspace{0.5ex}
		\multirow{2}{*}{East SST} &MAPE                   &13.03$\%$                   &18.63$\%$                   \\ \cline{2-4} 
		\vspace{0.5ex}
		\vspace{0.5ex}
		&Linear Regression       &y = $2.41*10^{-3}$x + 16.1             &y = $7.20*10^{-2}$x + 17.4          \\ \hline
		\vspace{0.5ex}
		\multirow{2}{*}{West SST} &MAPE                   &11.77$\%$                   &17.26$\%$                   \\ \cline{2-4}
		\vspace{0.5ex}
		\vspace{0.5ex}
		&Linear Regression  &y = $3.53*10^{-3}$x + 14.0  &y = $3.33*10^{-2}$x + 14.5  \\ \hline
		\vspace{0.5ex}
		\multirow{2}{*}{South SST} &MAPE                   &5.2$\%$                   &6.64$\%$                   \\ \cline{2-4} 
		\vspace{0.5ex}
		\vspace{0.5ex}
		&Linear Regression   &y = $1.83*10^{-3}$x + 18.3  &y = $3.35*10^{-2}$x + 19.0  \\ \hline
	\end{tabular}
	\caption{The sea surface temperatures (SST) in the East, West, and South Seas using ARIMA and Bayesian Structural Time Series (BSTS) models. The ARIMA and BSTS models exhibit particularly high accuracy in the South Sea, predicting a significant increase in SSTs across all three regions.}
	\label{tab:EWS SST ARIMA BSTS}
\end{table*}

The table \ref{tab:EWS SST ARIMA BSTS} provides a comprehensive analysis of the linear regression equations and the Mean Absolute Percentage Error (MAPE) for temperature predictions generated by both the ARIMA and Bayesian Structural Time Series (BSTS) models. In this comparison, the ARIMA model showed a slightly lower MAPE than the BSTS model; however, the difference was minimal and did not reach statistical significance. Nonetheless, the slope of the regression equation in the BSTS model tended to be larger than that in the ARIMA model, suggesting a potentially steeper predicted increase in temperature. It should be noted that both models displayed reasonable accuracy in predicting temperatures for the South Sea region.

\begin{table}[hbt!]
	\begin{tabular}{m{1.6cm} m{2.7cm} m{4cm}}
		\hline
		\vspace{0.5ex}
		        & Linear regression & Bayesian linear regression  \\ \hline \hline
		\vspace{0.5ex}
		East SSH           &\multirow{2}{*}{y = 0.242x + 17.1}  &  \multirow{2}{*}{y = 0.24$x_{-0.07}^{+0.08}$ + $17.08_{-1.37}^{+1.50}$} \\
		(Mukho)            &                       &                                              \\ \hline
		\vspace{0.5ex}
		West SSH         &\multirow{2}{*}{y = 0.279x + 457}  &  \multirow{2}{*}{y = 0.28$x_{-0.10}^{+0.10}$ + $457.07_{-2.17}^{+2.16}$}  \\ 
		(Incheon)          &                       &                                             \\ \hline
		\vspace{0.5ex}
		South SSH        &\multirow{2}{*}{y = 0.283x + 66.8} & \multirow{2}{*}{y = 0.28$x_{-0.08}^{+0.09}$ + $66.85_{-1.74}^{+1.79}$}  \\
		(Busan)            &                         &                                           \\ \hline
		\vspace{0.5ex}
	\end{tabular}
	\caption{The sea surface height (SSH) in the East, West, and South Seas using linear regression and Bayesian linear regression models (See Fig.\,\ref{fig:Bayesian dynesty}). In Bayesian linear regression, the + and - values denote the 1$\sigma$ intervals. The slopes and intercepts from both Linear Regression and Bayesian Linear Regression models show a close alignment.}
	\label{tab:EWS SSH LR BLR}
\end{table}

The table \ref{tab:EWS SSH LR BLR} presents a detailed comparison between the linear regression equation derived from the ARIMA model and the Bayesian linear regression equation formulated using the 'dynesty'\,\cite{J. S. Speagle_2020_dynesty} algorithm. The ARIMA model provides a straightforward linear regression equation based on historical data. In contrast, the Bayesian linear regression approach, utilizing 'dynesty', offers a more nuanced and probabilistic perspective. The predictions from the Bayesian model reveal a notable and observable increase in sea surface heights across the three seas. These findings highlight the capability of Bayesian linear regression to effectively identify and model intricate patterns and trends in sea surface heights, offering crucial insights for future climate and environmental evaluations.

\begin{table}[hbt!]
	\begin{tabular}{m{3.3cm} m{1.6cm} m{1.6cm} m{1.6cm}}
		\hline
		\vspace{0.5ex}
		  &1989&2003&2025  \\ \hline \hline
		\vspace{0.5ex}
		East SSH (Mukho)&18.5  &20.3  &\(25.1 \pm 2.86\)  \\ \hline
		\vspace{0.5ex}
		West SSH (Incheon)&459.5  &462.0  &\(466.7 \pm 1.75\)  \\ \hline
		\vspace{0.5ex}
		South SSH (Busan)&69.0  &71.3  &\(77.1 \pm 2.15\) \\ \hline
	\end{tabular}
	\caption{Actual sea surface height (SSH) values in centimeters (cm) for the years 1989 and 2003, alongside projected values for 2025 in the East, West, and South Seas. The forecasted values indicate a general increase in sea levels across all three regions.}
	\label{tab:EWS actual and predic SSH}
\end{table}

Table \ref{tab:EWS actual and predic SSH} represents the actual values in 1989 and 2003 and the predictive value in 2025. 
Tables \ref{tab:EWS SST ARIMA BSTS}, \ref{tab:EWS SSH LR BLR} and \ref{tab:EWS actual and predic SSH} elucidate the uptrends in SST and height across Korea using ARIMA model, BSTS model, Linear regression and Bayesian linear regression.
Given the premise that a 1 uptick in SST augments air humidity by 7\%, the chances of catastrophic events such as super typhoons surge proportionally\,\cite{M. Liu_2019}.
Rising SST not only elevate the sea's own temperature but also expand its volume. Such dynamics yield multiple consequences: an elevation in SSH, acidification of oceans, and a consequent reduction in marine biodiversity\,\cite{B. Talukder_2022}. 

Time series analysis provides a robust method to discern future trends based on past and present observations. One key distinction in this method is the differentiation of data into Stationary and Non-Stationary models.

\begin{enumerate}
	\item \textbf{Defining Stationarity and Non-Stationarity}: Stationary Model retains its statistical properties, such as mean or variance, irrespective of the time frame, meaning the data remains consistent over time. In contrast, Non-Stationary Model is a model that statistical properties evolve with time. To make meaningful future predictions using non-stationary data, it's crucial to transform it into a stationary model, as the forecast range for non-stationary data is potentially limitless. This transformation often involves either differencing the series or using a combination of differencing and natural logarithms.
	
	\item \textbf{Identifying Model Type with ACF and PACF}: Auto Correlation Function (ACF) measures autocorrelation based on time lags, accounting for values between all time points.
	Partial Auto Correlation Function (PACF) focuses only on the correlation between two distinct points in time, excluding intermediate intervals.
	
	\item \textbf{Classification of Stationary Models Using ACF and PACF}: In the realm of stationary data, we recognize three primary models. The tuple (p, d, q) describes the order of these models, where 'p' represents the autoregressive term, 'd' signifies the number of differences, and 'q' denotes the moving average.
	
	For data that doesn't align with any of these three stationary models, the ARIMA model, a non-stationary variant, may be applied. (p,q,d) can be seen through R program.
\end{enumerate}

\begin{table*}[hbt!]
	\begin{tabular}{m{3cm} m{7cm} m{7cm}}
		\hline
		\vspace{0.5ex}
		Staionary model          &ACF                   &PACF                   \\ \hline \hline
		\vspace{0.5ex}
		AR(p)           &\multirow{2}{*}{Exponentially decreasing curve or wave curve}  &  \multirow{2}{*}{Cutting from lag p} \\
		Auto Regressive            &                       &                                              \\ \hline
		\vspace{0.5ex}
		MA(q)          &\multirow{2}{*}{Cutting from lag q}  &  \multirow{2}{*}{Exponentially decreasing curve or wave curve}  \\ 
		Moving Average          &                       &                                             \\ \hline
		\vspace{0.5ex}
		ARMA(p,q)&Exponentially decreasing curve or wave curve                   &Exponentially decreasing curve or wave curve                   \\
		AR+MA&Exponentially decreasing from lag q                   &Exponentially decreasing from lag p                   \\ \hline
		\vspace{0.5ex}
	\end{tabular}
	\caption{Analysis of Auto Correlation Function (ACF) and Partial Auto Correlation Function (PACF). ACF examines autocorrelation across various time lags, while PACF assesses the direct correlation between specific time points, excluding intermediate lags. The tuple (p, d, q) defines the model order, where 'p' represents the autoregressive term, 'd' denotes the number of differences, and 'q' indicates the moving average component. For a detailed explanation of p, d, q, and the model, refer to Appendix B.}
\end{table*}

\section{Analyzing Temperature Trends with the ARIMA Model and Bayesian approach}
The ARIMA model, known for its sensitivity to rapid data fluctuations, is a suitable choice for many time series datasets, prompting our selection for analyzing temperature trends in Seoul.

We use the data from January 1980 to December 2021, and predict Seoul temperature from January 2022 onwards.

\begin{table}[hbt!]
    \centering
    \begin{tabular}{
        m{3.3cm} 
        >{\centering\arraybackslash}m{1.4cm} 
        >{\centering\arraybackslash}m{1.4cm} 
        >{\centering\arraybackslash}m{1.8cm}}
		\hline
		\vspace{0.5ex}
		  &1980&2021&2041  \\ \hline \hline
		\vspace{0.5ex}
		August Temp. (\(^\circ\mathrm{C}\))&22.8  &25.9  &\(26.63 \pm 1.45\)  \\ \hline
		\vspace{0.5ex}
		December Temp. (\(^\circ\mathrm{C}\))&-3.8  &0.6  &\(-0.14 \pm 1.45\)  \\ \hline
		\vspace{0.5ex}
	\end{tabular}
	\caption{Observed  temperatures in Seoul for the years 1980 and 2021, along with predicted values for 2041 using the ARIMA model. The predicted values include a ± range representing the 1$\sigma$ uncertainty.}
	\label{tab:ARIMA_Seoul}
\end{table}

Tables \ref{tab:ARIMA_Seoul} and \ref{tab:BSTS_Seoul} are related to Table \ref{tab:Seoul temp using ARIMA BSTS}.
Table \uppercase\expandafter{\romannumeral8} presents the actual and predicted temperatures in Seoul using the ARIMA model. 
Despite the fluctuations and variances in winter temperatures observed over different periods, the general long-term trend is unmistakably upward. 

\subsection*{Appendix C1: Bayesian Approaches in Time Series Analysis}

Historically, time series models like ARIMA have primarily relied on the frequentist approach. However, there has been a noticeable shift towards Bayesian methodologies in recent years. To understand the application of Bayesian Structural Time Series (BSTS) models, it is important to first become familiar with the foundational principles of structural time series models. These models decompose time series data into components such as trend and seasonality, providing a clear framework for analysis and interpretation.

This model can be split into two distinct equations:

\begin{enumerate}
	\item \textbf{Observation Equation}: This equation represents the observed data, \(y_t\), as a function of the latent variable, \(z_t\). Mathematically, it's given by:
	\begin{equation}
		y_t = \theta_t \cdot z_t + \epsilon_t
	\end{equation}
	
	\item \textbf{Transition Equation}: This equation delineates how the latent variable, \(z_t\), transitions or changes over time. It is given by:
	\begin{equation}
		z_{t+1} = \phi_t \cdot z_t + \omega_t \cdot \eta_t
	\end{equation}
\end{enumerate}

In these equations, \(\epsilon_t \sim N(0,\sigma^2_\epsilon)\) and \(\eta_t \sim N(0,\sigma^2_\eta)\) represent errors. The symbols \(\theta_t\), \(\phi_t\), and \(\omega_t\) act as structural parameters, with the outcomes of both equations primarily influenced by the latent variable, \(z_t\).

For a more simplified representation, we can configure the local level model by setting the observed data in place of the latent variable and normalizing the structural parameters to one:
\begin{equation}
	y_t = z_t + \epsilon_t
\end{equation}
\begin{equation}
	z_{t+1} = z_t + \eta_t
\end{equation}

To encapsulate the seasonality within a model, an additional term, \(s_t\), can be integrated into Equation (C3). Similarly, for regression characteristics, the term \(x_t\beta\) can be included, denoting the regressor and its corresponding coefficient.

For our analysis, we employed the Bayesian Structural Time Series (BSTS) package. This recently developed tool harnesses Bayesian principles to craft accurate time series models. Refer to references \,\cite{K. H. Brodersen_2015_Bayes, O. Poyser_2019_Bayes} for comprehensive information on BSTS.

\subsection*{Appendix C2: Bayesian Structural Time Series Analysis for Temperature Prediction in Seoul}

Remarkably, in Table \ref{tab:Seoul temp using ARIMA BSTS}, a MAPE of \(3.82\%\) is roughly twice as precise as the ARIMA-based time series analysis.
And the Bayesian Structural Time Series prediction slope was approximately 25 times steeper than the ARIMA prediction.

\begin{table}[hbt!]
    \centering
    \begin{tabular}{
        m{3.3cm} 
        >{\centering\arraybackslash}m{1.4cm} 
        >{\centering\arraybackslash}m{1.4cm} 
        >{\centering\arraybackslash}m{1.8cm}}
		\hline
		\vspace{0.5ex}
		   &1980&2021&2041  \\ \hline \hline
		\vspace{0.5ex}
		August Temp. (\(^\circ\mathrm{C}\))&22.8  &25.9  &\( 29.4^{+4.52}_{-4.71} \)  \\ \hline
		\vspace{0.5ex}
		December Temp. (\(^\circ\mathrm{C}\))&-3.8  &0.6  &\( 3.28^{+4.45}_{-6.24} \)  \\ \hline
		\vspace{0.5ex}
	\end{tabular}
	\caption{Same as Table \,\ref{tab:ARIMA_Seoul}, but using the BSTS model. The predicted temperature for Seoul in August 2041 is expected to approach approximately 30°C.}
	\label{tab:BSTS_Seoul}
\end{table}

Table \ref{tab:BSTS_Seoul} provides a detailed comparison of the actual and predicted temperatures in Seoul using the Bayesian Structural Time Series (BSTS) model. This table allows for a clear assessment of the model's predictive accuracy and the trends it forecasts. In contrast, Table \ref{tab:ARIMA_Seoul} presents temperature predictions for Seoul using the ARIMA model. When comparing the two tables, it is evident that the BSTS model predicts a significantly greater increase in temperatures compared to the ARIMA model. This difference underscores the improved sensitivity and forecasting capabilities of the BSTS model in identifying and predicting upward trends in temperature.

Summing up, the MAPE indicates the error rate disparity between actual and fitting values. While ARIMA's MAPE was \(7.98\%\), the BSTS model achieved a MAPE of \(3.82\%\). This infers the latter's double precision over the former. Consequently, the predicted linear regression slope intensifies from y=$4.29*10^{-4}$x + 13.0 in ARIMA to y=$1.06*10^{-2}$x + 13.03 in BSTS, with the more accurate BSTS predicting a more pronounced temperature rise.

\subsection*{Appendix C3: Bayesian Structural Time Series Analysis of Sea Temperatures}

Table \ref{tab:EWS SST ARIMA BSTS} illustrates the Bayesian structural time series analysis for the East Sea, West Sea, and South Sea.
Though the BSTS model displayed slightly higher errors compared to the prior ARIMA analysis, the difference wasn't substantial. 
In BSTS model, we represents a 1$\sigma$ uncertainty .

\begin{table}[hbt!]
	\begin{tabular}{m{2.2cm} m{3.5cm} m{2.5cm}}
		\hline
		\vspace{0.5ex}
		    &Date    &SST(\(^\circ\mathrm{C}\))     \\ \hline \hline
		\vspace{0.5ex}
		\multirow{3}{*}{East SST}&December 2002  &6.65     \\ \cline{2-3}
		\vspace{0.5ex}
		\vspace{0.5ex}
		&December 2012  &12.7    \\ \cline{2-3}
		\vspace{0.5ex}
		\vspace{0.5ex}
		&December 2033  &\( 20.1^{+5.68}_{-6.24} \)    \\ \hline \hline
		\vspace{0.5ex}
		\vspace{0.5ex}
		&December 2002  &7.89   \\ \cline{2-3} 
		\vspace{0.5ex}
		\vspace{0.5ex}
		West SST&December 2012  &9.16    \\ \cline{2-3}
		\vspace{0.5ex}
		\vspace{0.5ex}
		&December 2033  &\( 13.7^{+2.31}_{-2.71} \)    \\ \hline \hline
		\vspace{0.5ex}
		\vspace{0.5ex}
		&December 2002  &15.3   \\ \cline{2-3}
		\vspace{0.5ex}
		\vspace{0.5ex}
		South SST&December 2022  &16.2    \\ \cline{2-3}
		\vspace{0.5ex}
		\vspace{0.5ex}
		&December 2033  &\( 19.1^{+1.66}_{-2.05} \)    \\ \hline
		\vspace{0.5ex}
	\end{tabular}
	\caption{Observed  sea surface height (SSH) values for the years 2002 and 2012, along with predicted values for 2033 in the East Sea (Mukho), West Sea (Incheon), and South Sea (Busan) using the BSTS model. As shown in Fig. \ref{fig:annual_average_sea_surface_height}, the East Sea exhibits more pronounced fluctuations in actual SSH compared to the West and South Seas, which may be reflected in the larger 1$\sigma$ uncertainty.}
	\label{tab:BSTS_EWS_SST}
\end{table}

Table \ref{tab:BSTS_EWS_SST} is associated with Table \ref{tab:EWS SST ARIMA BSTS}.
Table \uppercase\expandafter{\romannumeral10} represents the actual and predicted sea surface temperatures for the East Sea, West Sea, and South Sea. 
The data indicate an increase in sea surface temperatures across all three regions.
The smaller MAPE value for the South Sea, as shown in Table \ref{tab:EWS SST ARIMA BSTS}, is reflected in the 1$\sigma$ uncertainty for this region.

The BSTS model predicts an increase in sea surface temperature, with projections ranging from 4°C to 14°C. In particular, the slope for the East Sea, as shown in Table \ref{tab:EWS SST ARIMA BSTS}, is approximately twice as steep as that for the West and South Seas. This indicates a much more pronounced rise in temperature in the East Sea. Overall, the BSTS analysis suggests significantly higher future sea surface temperatures when compared to the projections from the ARIMA analysis.

\subsection*{Appendix C3: ARIMA Model for Sea Surface Height}

In the selection of our time series model, we once again opted for the ARIMA model. This decision was driven by the model's applicability to diverse time series data and its ability to sensitively capture rapid fluctuations over time. For our analysis, we utilized the dataset which encompasses data from Mukho in the East Sea, Incheon in the West Sea, and Busan in the South Sea spanning the period from 1989 to 2021.

With the above foundation laid out, our subsequent steps involve predicting the alterations in sea surface height through Bayesian linear regression.

\section{Bayesian linear regression using dynesty}

Our analysis commenced with the examination of data collected from three key locations: Mukho in the East Sea, Incheon in the West Sea, and Busan in the South Sea. By integrating both actual and forecasted values, we derived a comprehensive linear regression equation. This equation serves to encapsulate the relationship between the observed data and the predicted trends, providing a robust framework for our temperature analysis across these regions. The specific linear regression equation is as follows:

\begin{figure*}
	\centering
	\includegraphics[width=15cm, height=10cm]{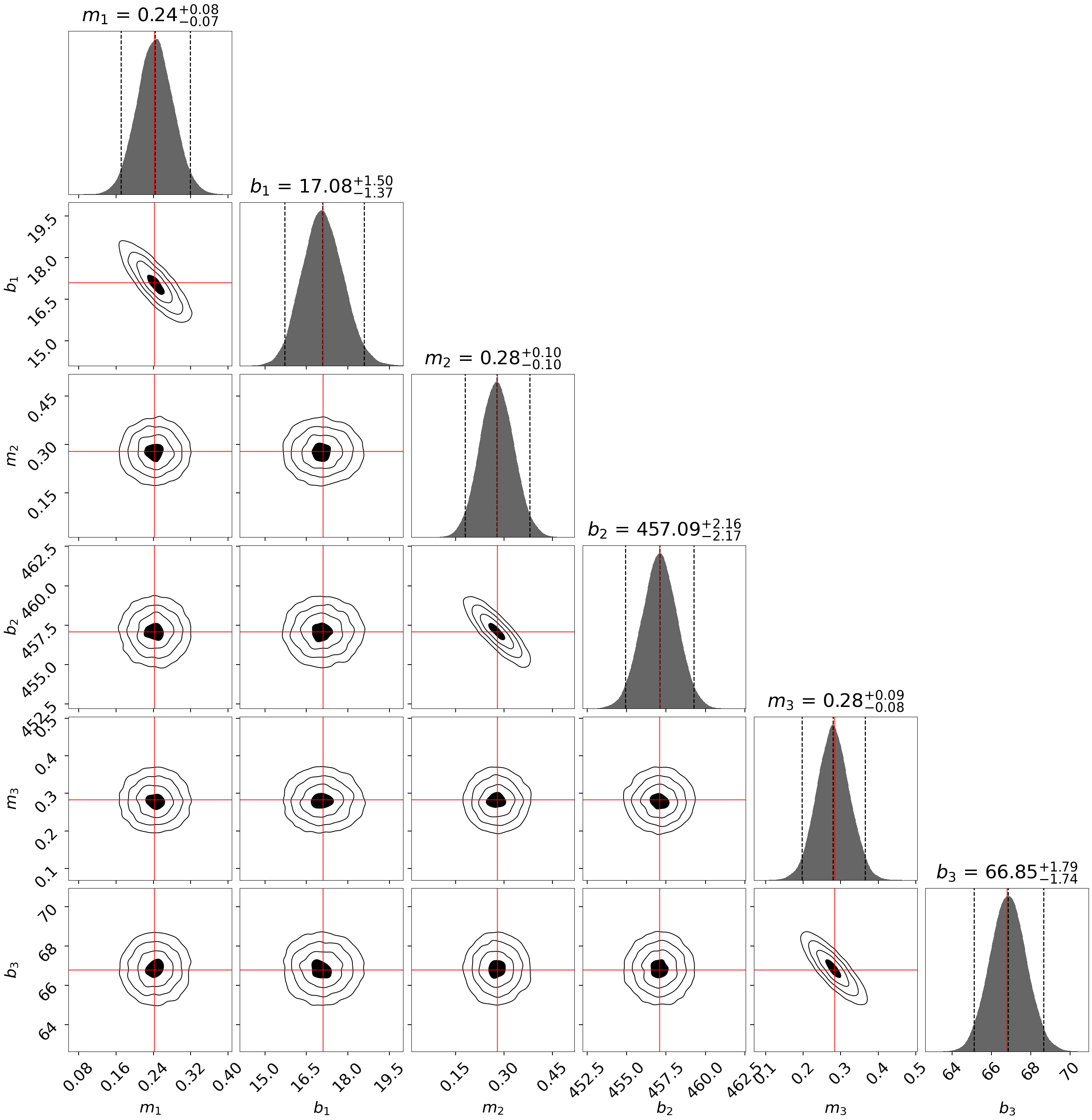}
	\caption{Bayesian linear regression analysis for Mukho, Incheon, and Busan using the dynesty analysis. The results are presented in Table \ref{tab:EWS SSH LR BLR}. In this graph, $m_1$ and $b_1$ represent the slope and intercept for the East Sea (Mukho), $m_2$ and $b_2$ correspond to the West Sea (Incheon), and $m_3$ and $b_3$ pertain to the South Sea (Busan). The ± signs next to each slope and intercept indicate the asymmetric 1$\sigma$ uncertainties, with the slopes and intercepts exhibiting a negative correlation.}
	\label{fig:Bayesian dynesty}
\end{figure*}

\begin{center}
East Sea(Mukho): $y$ = $0.24_{-0.07}^{+0.08}$ x + $17.08_{-1.37}^{+1.50}$
\end{center}

\begin{center}
West Sea(Incheon): $y$ = $0.28_{-0.10}^{+0.10}$ x + $457.09_{-2.17}^{+2.16}$
\end{center}

\begin{center}
South Sea(Busan): $y$ = $0.28_{-0.08}^{+0.09}$ x + $66.85_{-1.74}^{+1.79}$
\end{center}

Refer to Fig.\,\ref{fig:Bayesian dynesty} for the outcomes of the `dynesty` analysis based on the data. The East Sea corresponds to $m_1$ and $b_1$, the West Sea has $m_2$ and $b_2$, and the South Sea has $m_3$ and $b_3$. The following provides an example for the East Sea in Fig.\,\ref{fig:Bayesian dynesty}.
\begin{itemize}
	\item The mean slope $m_1$ was 0.24 with an asymmetric 1$\sigma$ deviation of +0.08 and -0.07.
	\item The mean intercept $b_1$ was 17.08, with an asymmetric 1$\sigma$ deviation of +1.50 and -1.37. 
	\item Both the distributions approximate Gaussian distributions, aligning closely with the linear regression equation.
	\item The graph depicts a negative correlation between the two variables; as the slope $m_1$ ascends with a constant intercept $b_1$.
\end{itemize}

Figure\,\ref{fig:Bayesian dynesty} demonstrate the `dynesty` outcomes for this dataset. The graphical structure remains consistent with our previous examples. Conclusively, a comparison between `dynesty` results and ARIMA time series findings showcases a high degree of alignment. Both methodologies, time series analysis and Bayesian linear regression, concur in predicting the sea surface rise across East Sea (Mukho), West Sea (Incheon), and South Sea (Busan). Refer to \,\cite{J. S. Speagle_2020_dynesty} for comprehensive information on the 'dynesty' analysis.

\end{document}